\documentclass[aps,pra, twocolumn, noshowpacs, floatfix]{revtex4}

\usepackage{amsfonts}
\usepackage{amssymb}
\usepackage{graphicx}
\usepackage{subfigure}
\usepackage{amsmath}
\usepackage[english]{babel}
\usepackage{color}
\usepackage{epstopdf}
\usepackage{footnote}

\begin{document}

\title{Exciton Dynamics and Time-Resolved Fluorescence in Nanocavity-Integrated Monolayers of Transition-Metal Dichalcogenides}

\author{Kewei Sun$^{1}$, Kaijun Shen$^{2}$, Maxim F. Gelin$^{1}$ and Yang Zhao$^{2}$\footnote{Electronic address:~\url{YZhao@ntu.edu.sg}}}
\affiliation{$^{1}$\mbox{School of Science, Hangzhou Dianzi University, Hangzhou 310018, China}\\
$^{2}$\mbox{School of Materials Science and Engineering, Nanyang Technological University, Singapore 639798, Singapore} \\
}

\begin{abstract}
	We have developed an ab-initio-based fully-quantum numerically-accurate methodology for the simulation of the exciton dynamics and time- and frequency-resolved fluorescence spectra of the  cavity-controlled two-dimensional materials at finite temperature  and applied this methodology to the single-layer WSe$_2$ system.  This allowed us to establish dynamical and spectroscopic signatures of the polaronic and polaritonic effects  as well as uncover their characteristic timescales in the relevant range of temperatures.
\end{abstract}
\date{\today}
\maketitle

\section{TOC Graphic}
\includegraphics[scale=0.8,trim=0 0 0 0]{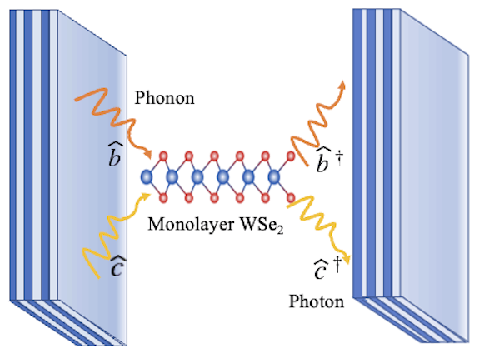}

Two-dimensional (2D) materials have shown fascinating electrical, mechanical, optical, and spintronic properties~\cite{Jin,Berghauser,Xia}. The development of valley electronics is inseparable from the study of 2D atomic layered materials. Early graphene was confirmed to have optical and electrical properties dependent on valleys, defined as the extreme points of the Bloch electron bands, by artificially breaking the symmetry of spatial inversion~\cite{Xiao,Yao}. Unlike graphene, the crystal structure of single-layer transition metal dichalcogenides (TMDs) does not have spatial inversion symmetry, leading to a nonzero Berry curvature. There is strong coupling between valleys and spins, and the polarization of a specific valley can be achieved by optical or electromagnetic injection of spins~\cite{Xiao1,Xie,Ye}. Properties of valley electronic materials depend on the atomic structure. With the decrease of the number of atomic layers, for example, electronic structures of TMDs undergo a transition from indirect semiconductors to direct semiconductors~\cite{Mak,Splendiani}. The difference in symmetry between single-layer and multi-layer TMDs materials also provides a way to artificially modulate the valley properties. Therefore, TMDs materials have become an important venue for studying valley effects and constructing valley devices~\cite{Xia,Yu}.
{Furthermore, as the surface of monolayer TMDs is free of dangling bonds and the interaction between layers is dominated by the van der Waals force, a heterostructure can be constructed by directly assembling single-layered TMDs vertically without resorting to lattice matching. TMDs-based heterostructures provide more freedom to engineer the optical and the electronic properties of quantum materials than in monolayer TMDs~\cite{Rivera,JinC,JiangY}.}

In single-layered TMDs, the bright spin-valley excitonic states located at the $\rm K$ or the $\rm K^{\prime}$ valleys can be optically excited by circularly polarized light, thus acquiring a valley degree of freedom~\cite{Berghauser,Ubrig}. In tungsten-based monolayer materials,
it has been verified by low-temperature photoluminescence (PL) emission that the intervalley excitons, i.e., the momentum-dark exciton states, have lower energy than that of the optically bright excitons~\cite{Brem}. Thus the momentum-dark excitons are expected to have strong influences over the optical and the electronic properties of TMDs.
{Moreover, many 2D materials are also known to interact strongly with light. The interaction must be further enhanced to reach the standard of actual use, for instance, TMDs-based nanolasers~\cite{Xia}. In general, two methods, integrating the 2D materials with optical cavities or using their intrinsic polaritonic resonances~\cite{Liu,Koppens}, are developed to achieve this goal. Cavity-controlled, single-layered TMDs can be realized by their direct integration with onchip, planar nanophotonic cavities, which makes the realization of compact devices possible~\cite{Ross}. The excitons in semiconducting TMDs couple with the photonic modes in nanocavities, leading to a promising platform for engineering novel light-matter interactions~\cite{Walther,Ryou}.}

PL is in general an important technique to probe intriguing quantum phenomena and many-body correlations in TMDs materials \cite{Kira}. At lower temperatures, PL emission points to the existence of bound excitons such as trions, biexcitons, and trapped excitons~\cite{Brem,Mak1,Ye1,Huang}.
The relaxation mechanism is crucial for intrinsic properties of TMD materials such as the coherence lifetime and the thermalization of excitons. PL emission exhibits strong phonon-assisted features which arise from a momentum indirect electron-hole pair recombination~\cite{Brem}. {Also, in the cavity-controlled 2D materials} the exciton-cavity and exciton-phonon interactions together lead to an asymmetric PL lineshape in the form of phonon sidebands, i.e., so-called cavity-coupled PL emission~\cite{Ross}. In this hybrid system, there exists extraordinarily rich many-body processes, such as the phonon-assisted intervalley population transfers, the indirect PL signal, cavity enhanced electroluminescence, etc. In addition, time-resolved fluorescence (TRF) spectroscopy~\cite{Hyeon,Ahn,Nakamura}, as a more general concept, is a technique to monitor interactions between molecules and nuclear motion that occurs in the short periods. It can be used to track the time after the excitation event, and the time delay allows one  to probe the relaxation processes in the excited states. {By virtue of the TRF technique}, we can have a more comprehensive understanding of population transfer between different excitonic states and the accompanying phonon-assisted recombination process.

However, most of the relevant theoretical studies are based on the perturbative approaches, for instance, quantum master equation method~\cite{Ross} and the cluster expansion method with the truncation scheme~\cite{Brem} (see Ref. \cite{Kira} for a review),
which fail in the case of strong light-matter interaction or strong exciton-phonon coupling.
Furthermore,  the temperature-dependent  shifts of PL spectra  are usually described by the semiempirical Varshni equation~\cite{Var,Shen}.
In recent years, a powerful and numerically-accurate technique, namely, the multiple Davydov Ansatz (DA) method, has been developed by Zhao and coworkers \cite{Zhao} and applied to a variety of multidimensional problems to scrutinize Landau-Zener transitions~\cite{WangL,ZhengF,HuangZ, Zheng2}, polariton dynamics in cavity-assistant singlet fission~\cite{Sun1,Sun2}, exciton dynamics in biological light-harvesting complexes~\cite{LP_JCP2015,LPJPCL2018,Fulu}, and ultrafast dynamics at conical intersections~\cite{skw1,skw2,Sun3}.  Furthermore,  the multiple DA method has been extended to finite temperatures \cite{Zhao}. In the present work, we  employ the  multiple DA approach combined with the thermo field dynamics (TFD) technique~\cite{Chen,Borrelli,Borrelli1} to give a fully microscopic and numerically accurate methodology
for the simulation of dynamic and spectroscopic  responses of cavity-controlled single-layered TMD materials at final temperature and apply it to scrutinize the exciton dynamics and TRF spectra of the single-layered cavity-controlled  WSe$_2$ system.

The  Hamiltonian characterizing the exciton-polariton dynamics in the undoped monolayer cavity-controlled WSe$_2$ system reads~\cite{Brem,Ross,Ivanov,Katsch}
\begin{eqnarray}
\label{H}
H&=&\sum_{(i\bf Q_{\parallel})}E_{(i\bf Q_{\parallel})}a_{(i\bf Q_{\parallel})}^{\dagger}a_{(i\bf Q_{\parallel})}+\hbar\omega_{\sigma_{+}}c_{\sigma_{+}}^{\dagger}c_{\sigma_{+}}\nonumber\\
&&+M_{\sigma_{+}}[c_{\sigma_{+}}^{\dagger}a_{(1\bf \Gamma)}+c_{\sigma_{+}}a_{(1\bf \Gamma)}^{\dagger}]+\sum_{\alpha \bf q_{\parallel}}\hbar\Omega_{\alpha,\bf q_{\parallel}}b_{\alpha,\bf q_{\parallel}}^{\dagger}b_{\alpha,\bf q_{\parallel}}\nonumber\\
&&+\sum_{ij\alpha \bf Q_{\parallel} \bf q_{\parallel}}D^{ij}_{\alpha \bf q_{\parallel}}a^{\dagger}_{(j\bf Q_{\parallel}+\bf q_{\parallel})}a_{(i\bf Q_{\parallel})}(b^{\dagger}_{\alpha,-\bf q_{\parallel}}+b_{\alpha,\bf q_{\parallel}}).
\end{eqnarray}
Here operators $a_{(i\bf Q_{\parallel})}^{(\dagger)}$, $c_{\sigma_{+}}^{(\dagger)}$, $b_{\alpha,\bf q_{\parallel}}^{(\dagger)}$ ($a_{(i\bf Q_{\parallel})}$, $c_{\sigma_{+}}$, $b_{\alpha,\bf q_{\parallel}}$) create (annihilate), respectively,  excitonic states $(i\bf Q_{\parallel})$, photonic circularly-polarized mode  $\sigma_{+}$ and phonon modes $\alpha$ with momentum $\bf q_{\parallel}$ and frequency $\Omega_{\alpha,\bf q_{\parallel}}$. Energies $E_{(i\bf Q_{\parallel})}$ determine the excitonic band structure, the symbol $\parallel$ denotes the component of the momentum parallel to the monolayer plane,  $M_{\sigma_{+}}$ is the exciton-photon coupling coefficient and $D^{ij}_{\alpha \bf q_{\parallel}}$ are the exciton-phonon coupling matrix elements.
{In this work, the cavity dissipation is neglected to better clarify the temperature effects on exciton dynamics. In fact, the ultrahigh-Q (Q$\sim$750000) PhCnB cavities based on a five-hole taper design have been reported in Ref.~\cite{Deotare}, which rationalizes our assumption.}

\begin{figure}[tbp]
	\centering
	\includegraphics[scale=0.4,trim=20 0 0 0]{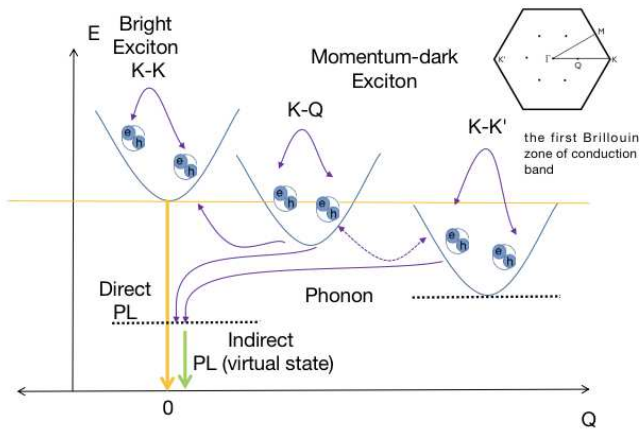} \\
	\caption{Schematic illustration of the formation of direct and indirect PL signals. The momentum-dark excitons $\rm KQ$ and $\rm KK^{\prime}$ can decay via radiative recombination by emitting or absorbing a phonon, which contributes to the indirect PL signal. The upper right inset shows the conduction electrons in the different valleys in the first Brillouin zone.}
	\label{Fig0}
\end{figure}

 The PL and TRF signals are mainly determined by the $\rm KK$, $\rm KQ$ and $\rm KK^{\prime}$ excitons with the momenta $1\bf\Gamma$, $2\bf\Lambda$ and $3\bf K$, while other higher-lying excitonic states are irrelevant at temperatures lower than room temperature. The intervalley excitons $\rm KQ$ and $\rm KK^{\prime}$ are optically dark and are responsible for the indirect phonon-assistant PL, while the intravalley exciton $\rm KK$ is optically bright and produces direct PL. The microscopic mechanisms behind direct and indirect PL are elucidated by Fig.~\ref{Fig0}.

In the present work, the exciton energies of the WSe$_2$ system are fixed at $E_{(1\bf\Gamma)}=1.724~\rm eV$, $E_{(2\bf\Lambda)}=1.69~\rm eV$, and $E_{(3\bf K)}=1.678~\rm eV$. To enhance the cavity-induced effects, the photonic mode is taken in resonance with the optically-bright $\rm KK$ exciton, $\hbar \omega_{\sigma_{+}} = 1.724~\rm eV$. The  exciton-photon coupling strength $M_{\sigma_{+}}$  can be adjusted in the range of $4~\rm meV\sim 14~\rm meV$ in the SiN ring resonator or photonic crystal cavity~\cite{Ross}, and the value of $M_{\sigma_{+}}=12~\rm meV$ is adopted in our calculations.
The frequencies of the phonon modes are listed in Table \ref{TT1} (cf. Ref.~\cite{Brem,Jin}). Both longitudinal acoustic (LA) and transverse acoustic (TA) modes show linear dispersion in the long wavelength limit and their frequencies are thus equal to zero for $q=0$. The three optical modes which can be strongly coupled to  electrons are the homopolar ($\rm A_1$, out-of-plane vibrations), the longitudinal (LO) and the transversal (TO) modes.

 The exciton-phonon coupling coefficients $D^{ij}_{\alpha \bf q_{\parallel}}$ are determined by the electron-hole-phonon scattering. Adopting the  density functional theory formalism (mean effective deformation potential approximation) and describing the exciton wave function in the momentum space, the coefficients can be evaluated as~\cite{Selig1,Brem2}
\begin{eqnarray}
	\label{D}
	D^{ij}_{\alpha \bf q_{\parallel}}=\sum_{\bf k}\Phi_{\bf k}^{i*}(\Phi_{\bf k+\mu\bf q_{\parallel}}^{j}g_{\alpha\bf q_{\parallel}}^{c{\bf k}(ij)}-\Phi_{\bf k-\nu\bf q_{\parallel}}^{j}g_{\alpha\bf q_{\parallel}}^{v{\bf k}(ij)}).
\end{eqnarray}
Here $\Phi_{\bf k}^{i}$ are the exciton wavefunctions,
\begin{eqnarray}
	\label{g}
	g_{\alpha\bf q_{\parallel}}^{c/v{\bf k}(ij)}=\sqrt{\frac{\hbar}{2\mathcal{M}\Omega_{\alpha,\bf q_{\parallel}}}}\langle j,{\bf k+q}_{\parallel}|\Delta V_{\bf q_{\parallel}}^{\alpha}|i,\bf k\rangle,
\end{eqnarray}
are the electron-phonon matrix elements in conduction/valence band, $\alpha$ and $\bf q_{\parallel}$ specify the scattered phonon mode,
$\mathcal{M}$ is the total mass of the atoms in the unit cell, and $|i,\bf k\rangle$ is the Bloch eigenstate with the wave vector
$\bf k$ and excitonic band index $i$. The perturbing potential $\Delta V_{\bf q_{\parallel}}^{\alpha}$ is obtained by the density-functional perturbation theory. The acoustic deformation potential, i.e. the first-order deformation potential, is given by the expression $\langle j,{\bf k+q}_{\parallel}|\Delta V_{\bf q_{\parallel}}^{\alpha}|i,\bf k\rangle/|\bf q_{\parallel}|$. It can be shown that the acoustic phonon at $\bf \Gamma$ does not contribute to electron-phonon coupling. The optical deformation potential, i.e. the zero-order deformation potential, is defined as $\langle j,{\bf k+q}_{\parallel}|\Delta V_{\bf q_{\parallel}}^{\alpha}|i,\bf k\rangle$~\cite{Li} and the parameters specifying this potential can be found in Ref.~\cite{Jin}.
To calculate exciton-phonon coupling strengths, we associate $\Phi_{\bf k}^{i}$ with the $1s$ eigenfunction of the Wannier equation in the momentum space~\cite{Berghauser},
\begin{eqnarray}
	\Phi_{\bf k}^{1s}=\frac{(2a_{\rm ex}\hbar)^{\frac{3}{2}}\hbar}{\pi(\hbar^2a_{\rm ex}^2k^2+\hbar^2)^2}
\end{eqnarray}
where $a_{\rm ex}=\epsilon_rm_e/\bar{\mathcal{M}}a_H$ is the excitonic radius, $\epsilon_r=4.5$ is the relative permittivity of HBN-encapsulated TMD monolayer~\cite{Brem}, $m_e$ is the electron mass, $a_H$ is the Bohr radius, and $\bar{\mathcal{M}}$ is the reduced mass of the electron-hole pair.

\begin{table}
\caption{Frequencies (in units of meV) of the phonon modes.}
\begin{center}

\begin{tabular}{p{2cm}p{2cm}p{2cm}p{2cm}p{2cm}}
\hline
\hline
$\rm Mode$ & $\bf \Gamma$ & $\bf \Lambda $ & $\bf K $ \\
\hline
$\rm TA$ & 0 & 11.6 & 15.6 \\
$\rm LA$ & 0 & 14.3  & 18 \\
$\rm TO$ & 30.5 & 27.3 & 26.7\\
$\rm LO$ & 30.8 & 32.5 & 31.5\\
$\rm A_1$ & 30.8 & 30.4 & 31 \\
\hline
\hline
\end{tabular}
\end{center}
\label{TT1}

\end{table}

This finalizes the construction of the microscopic model of the WSe$_2$ monolayer which is based on the ab initio input parameters such as the electronic band structure, phonon dispersion, and electron(hole)-phonon coupling strengths.{ All relevant ab-initio parameters can be found in Section $S5$ of the Supporting information.} The holes located in the vicinity of the K point are not included in the model since the corresponding states are $\rm KK^{\prime}$ symmetric. The spin-forbidden dark states requiring spin-flip processes are also neglected since they do not contribute to the system dynamics on the ultrafast timescale of interest. The developed model realistically describes the coherent dynamics of excitons, phonons and photons, as well as accounts for environmental dephasing~\cite{Brem}. The model Hamiltonian involves $23$ phonon modes, three excitonic states (KK, KQ, $\rm KK^{\prime}$), and a single photonic mode.  The total number of the phonon modes exceed 13 (see Table \ref{TT1}) due to the following two reasons. First,  the  $\pm$ modes $b_{\alpha,\pm \bf q_{\parallel}}$ with $\alpha=\rm KQ, KK^{\prime}$ have to be considered. Second, the total number of modes is doubled due to the introduction of the tilde modes $\tilde{b}_{\alpha,\bf q_{\parallel}}$ which are responsible for  temperature effects in the TFD framework \cite{Chen,Borrelli}.
It is essential that the potential energy surfaces of the three excitonic states  and a single photonic state cross each other via multidimensional conical intersections \cite{ConicalIntersections} shaped by 23 coupling modes. These conical intersections are responsible for the ultrafast exciton dynamics in the WSe$_2$ monolayer.

Having established and parameterized the model Hamiltonian, we combined the multi-$\rm D_{2}$ DA method \cite{Zhao} with the TFD machinery~\cite{Chen,Borrelli} to accurately simulate the exciton dynamics and the TRF spectra of the cavity-controlled WSe$_2$ monolayer at finite temperatures.
The underlying equations can be found in Supporting Information.  The simulations require evaluation of the system dynamics in the singly-excited excitonic manifold only, i.e. for  $\langle N_{\rm ex} \rangle = 1$. In the present model, the number of excitations is conserved because the number operator $N_{\rm ex}=c_{\sigma_{+}}^{\dagger}c_{\sigma_{+}}+\sum_{(iQ_{\parallel})}a_{(iQ_{\parallel})}^{\dagger}a_{(iQ_{\parallel})}$ commutes with the Hamiltonian $H$.  A multiplicity of 48 is used for obtaining convergent results at all considered temperatures.

\begin{figure*}[tbp]
\centering
\subfigure[]{
\includegraphics[scale=0.25,trim=80 0 80 0]{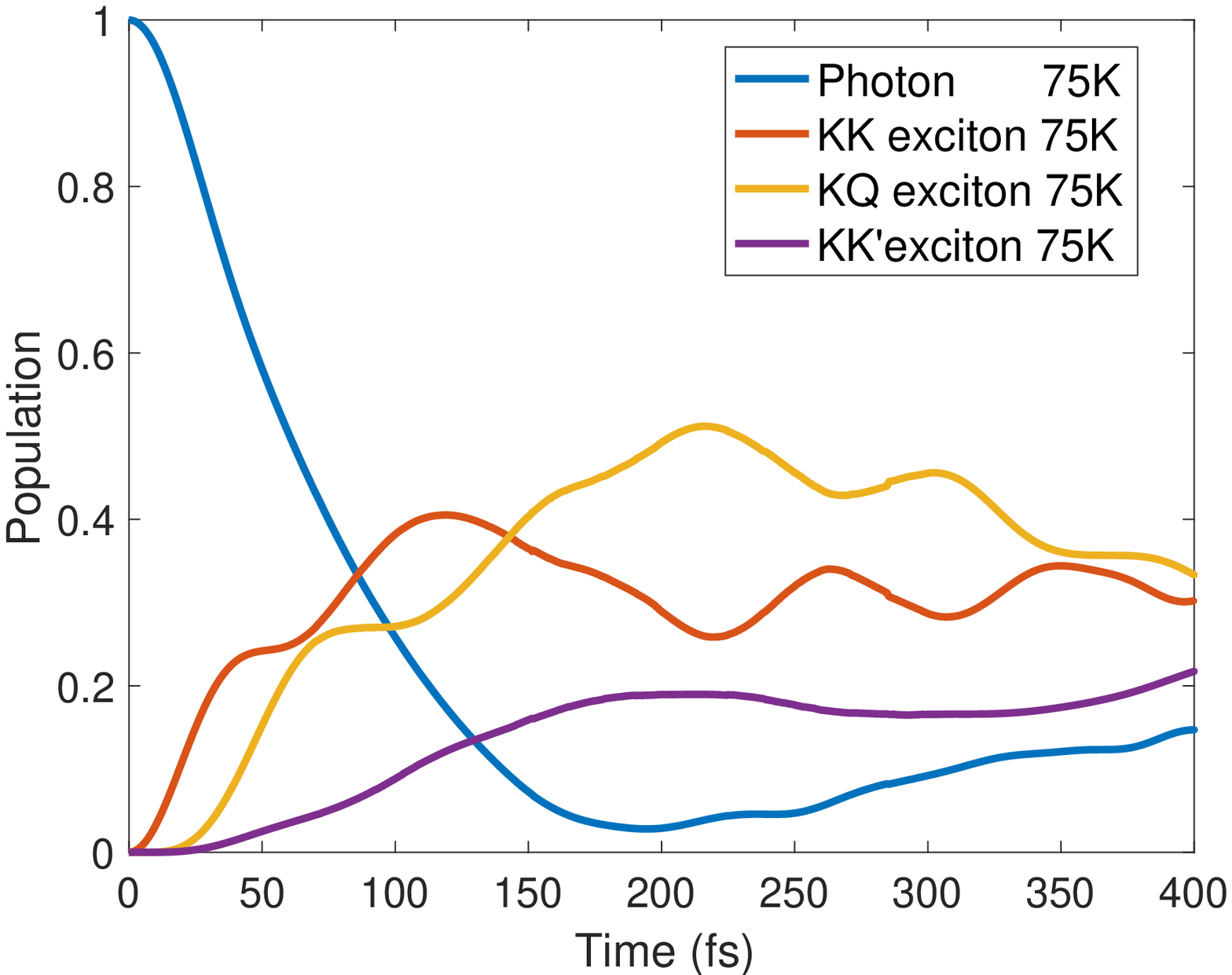}
}
\quad
\subfigure[]{
\includegraphics[scale=0.25,trim=30 0 80 0]{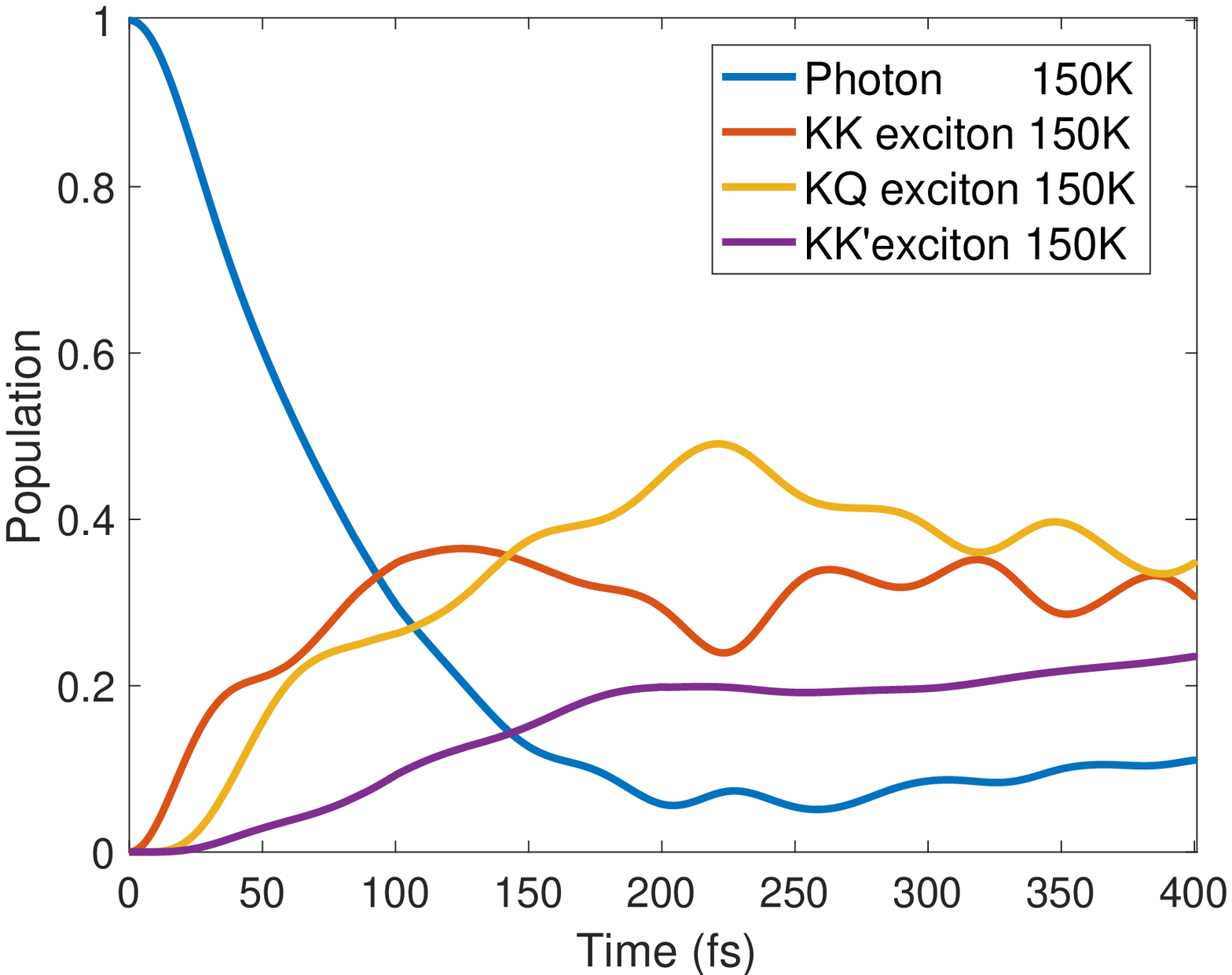}
}
\quad
\subfigure[]{
\includegraphics[scale=0.25,trim=30 0 80 0]{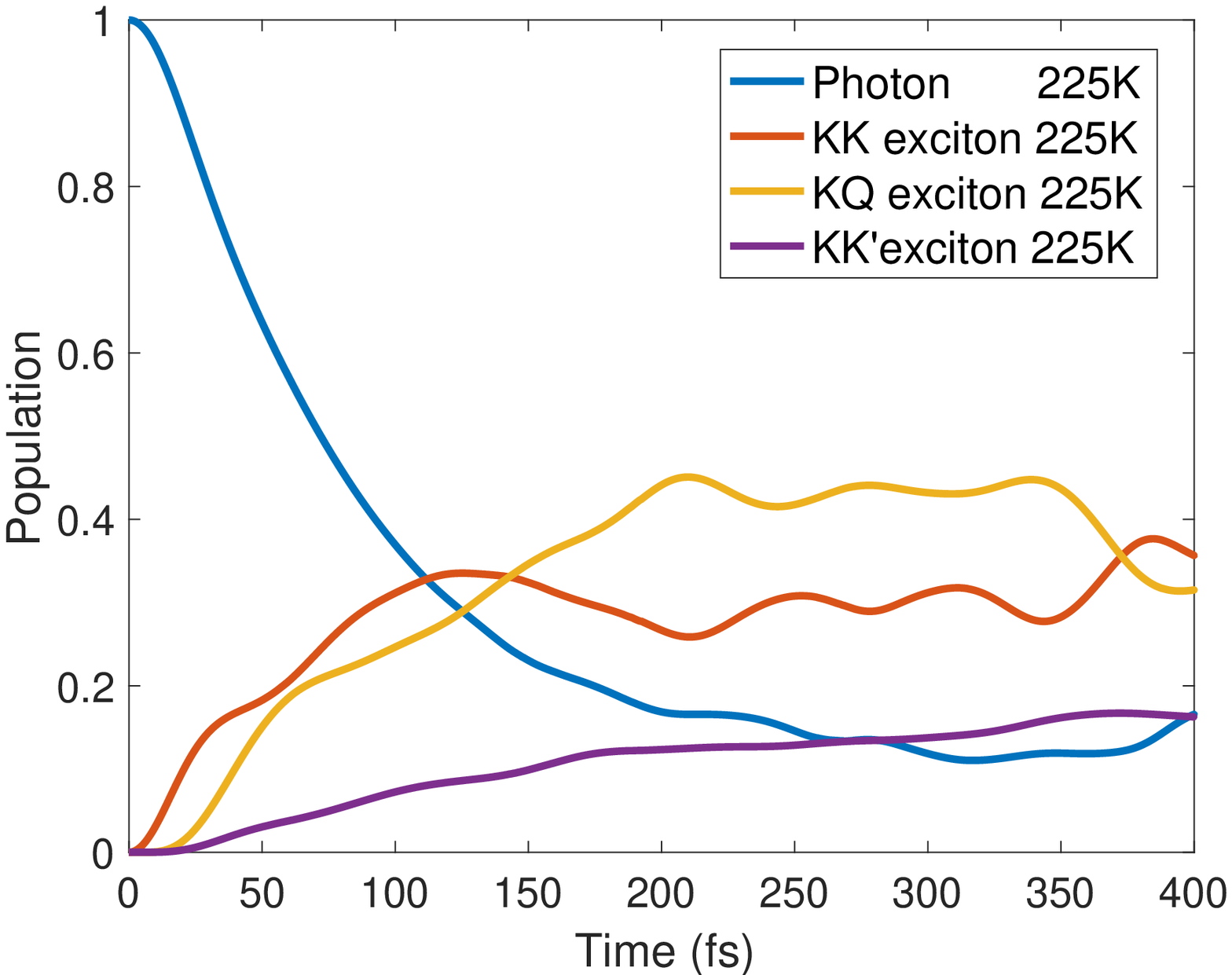}
}
\quad
\subfigure[]{
\includegraphics[scale=0.25,trim=30 0 80 0]{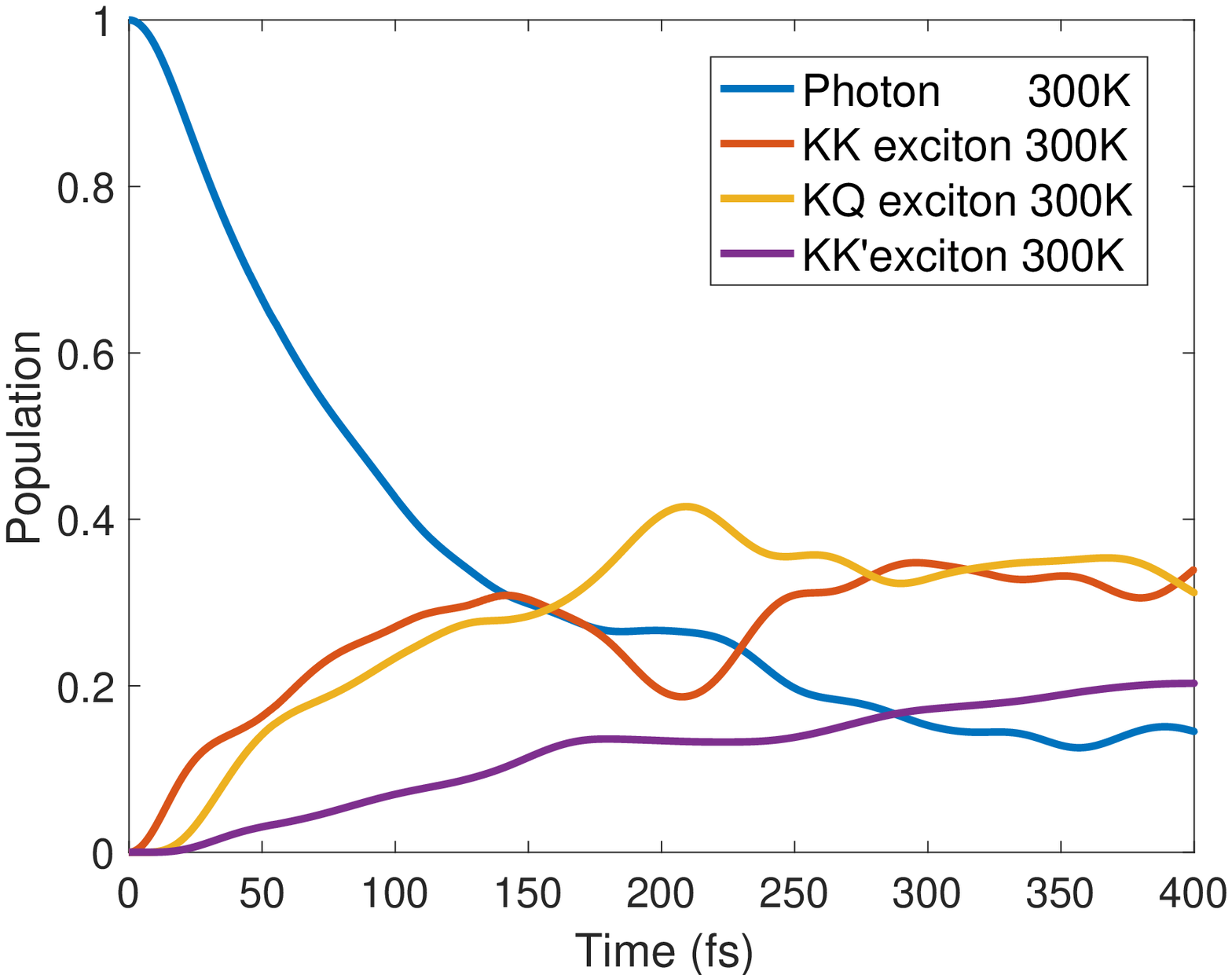}
}\\
\subfigure[]{
\includegraphics[scale=0.25,trim=80 0 80 0]{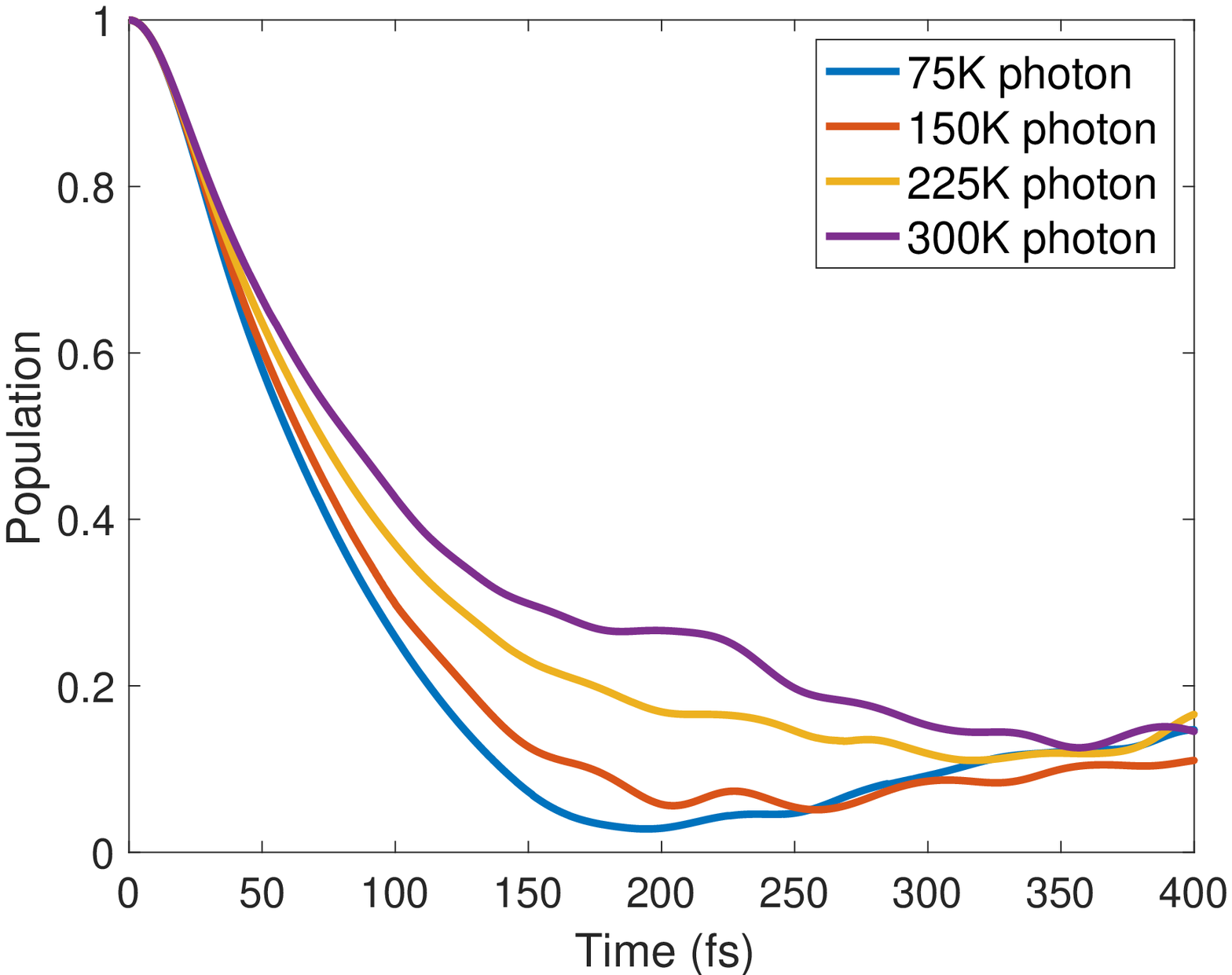}
}
\quad
\subfigure[]{
\includegraphics[scale=0.25,trim=30 0 80 0]{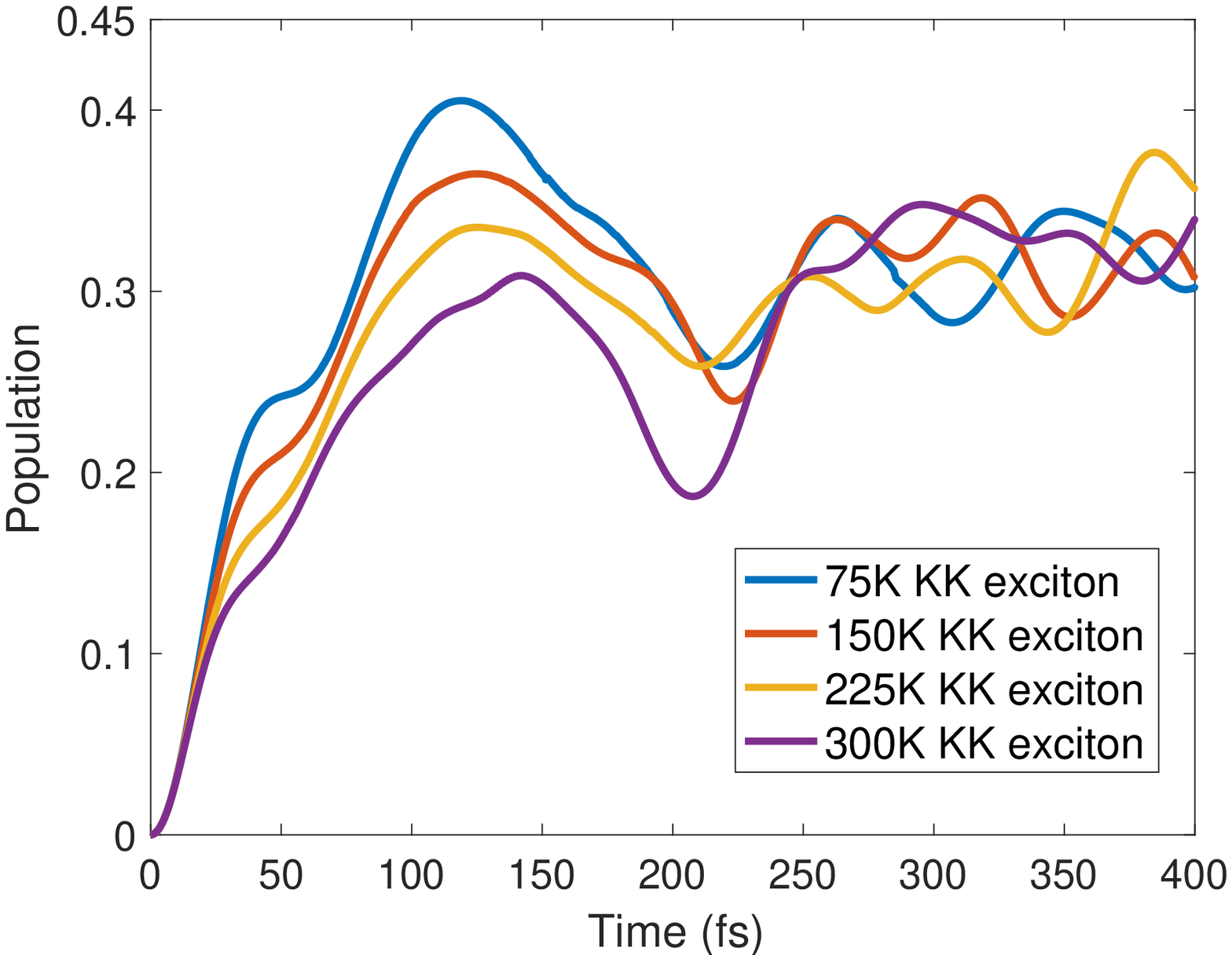}
}
\quad
\subfigure[]{
\includegraphics[scale=0.25,trim=30 0 80 0]{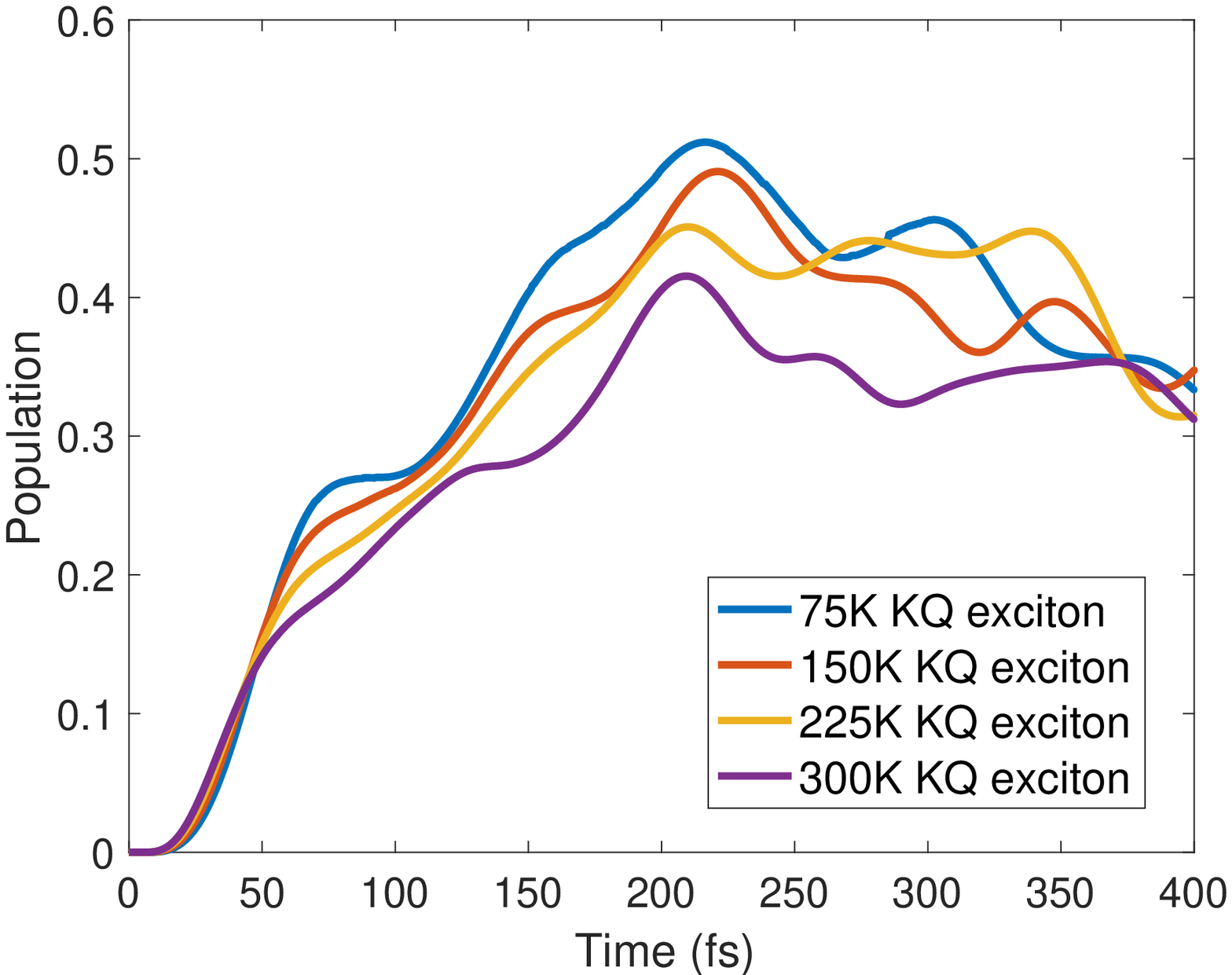}
}
\quad
\subfigure[]{
\includegraphics[scale=0.25,trim=30 0 80 0]{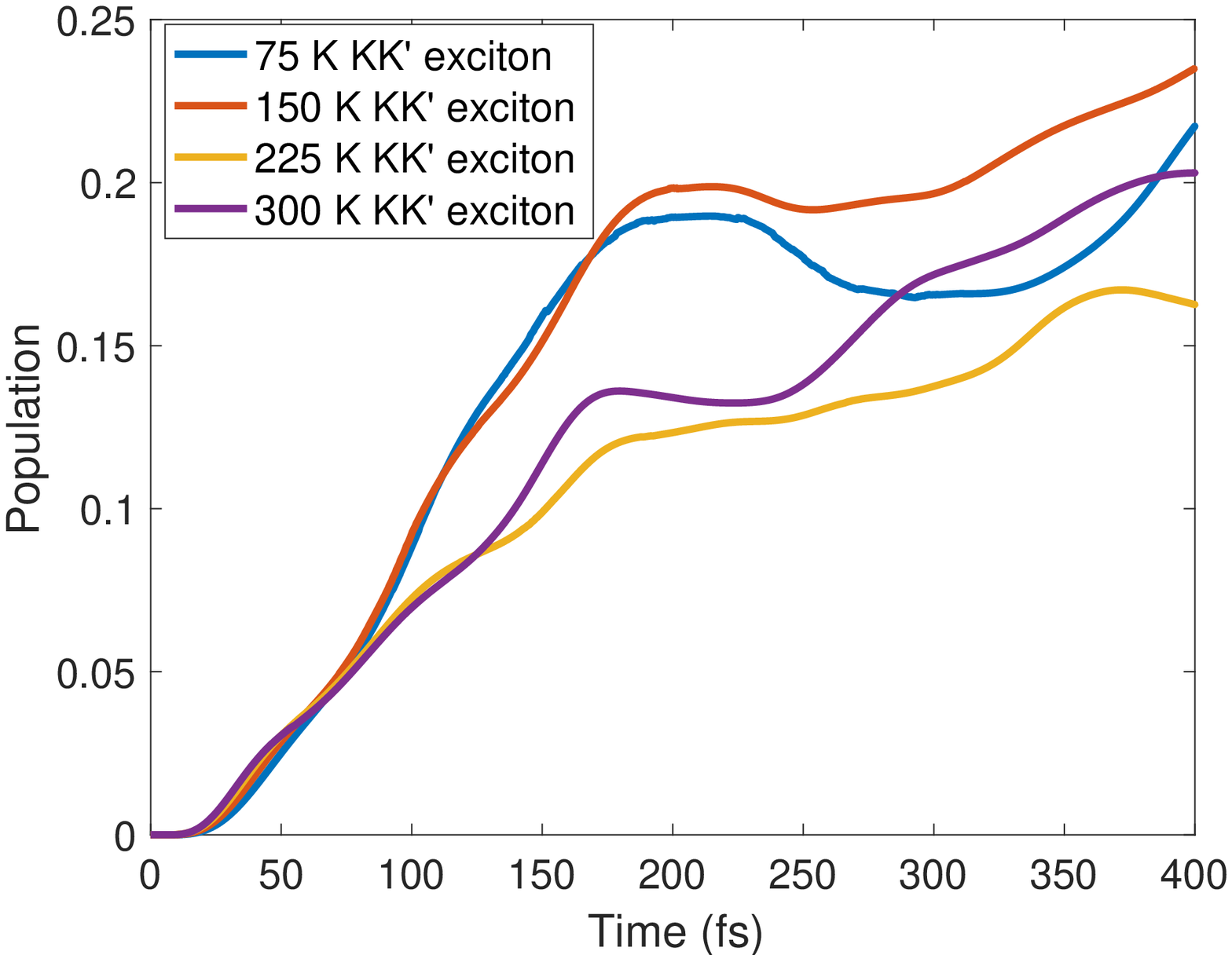}
}\\
\caption{Upper  panels: Populations of the photononic and excitonic modes of the WSe$_2$ system  at temperatures (a) 75 K, (b) 150 K, (c) 225 K and (d) 300 K. Lower panels:  Population of the  photonic (e) and excitonic KK (f), KQ (g), $\rm{KK^{\prime}}$  (h)  modes at four temperatures indicated in the panels.  In panel (h), the dashed lines indicate different stages of the population evolution.}
\label{Fig1}
\end{figure*}

Population evolutions of three excitonic states KK, KQ, $\rm{KK^{\prime}}$ and the photon mode are shown in the upper panels of Fig.~\ref{Fig1} for temperatures 75 K (a), 150 K (b), 225 K (c) and 300 K (d). The photon-mode populations first drop to their minima on the timescale from 200 fs (a) to 400 fs (d) and then exhibit low-amplitude oscillations around a certain averaged value. As temperature increases, positions of the minima move towards longer times while the minimal population values increase. Since only the photonic mode is initially  excited, excitonic populations increase at $t<150$ fs: The KK-mode exhibits the fastest growth and it is followed by the KQ and $\rm{KK^{\prime}}$ modes. This mirrors the predominantly sequential population transfer. At longer times, this rule ceases and the populations exhibit low-amplitude oscillations on their way to attain the corresponding limiting  values. However, the $\rm{KK^{\prime}}$-mode population is always smaller than populations of the KK and KQ modes, at least up to 1 ps. Incidentally, the KK and KQ populations are almost mirror images of each other for $150<t<350$ fs.

The lower panels of  Fig.~\ref{Fig1} display populations of the photonic (e) and excitonic KK (f), KQ (g), $\rm{KK^{\prime}}$ (h) modes at four temperatures indicated on the panels. At $t< 250$  fs, the photonic mode population increases and the bright-exciton population decreases with temperature. This can be understood by invoking the TFD framework, in which exciton-phonon coupling strengths of the physical ($b_{\alpha, \bf q_{\parallel}}$) and tilde ($\tilde{b}_{\alpha, \bf q_{\parallel}}$) phonon modes are proportional, respectively,  to  the temperature-dependent factors $\cosh(\theta_{\alpha, \bf q_{\parallel}})$ and  $\sinh(\theta_{\alpha, \bf q_{\parallel}})$, where $\theta_{\alpha, \bf q_{\parallel}}=\rm arctanh(\exp [-\hbar \beta \Omega_{\alpha, \bf q_{\parallel}}/2])$ and  $\beta = k_BT$ \cite{Chen,Borrelli}. The total exciton-phonon coupling strength, which is proportional to $\cosh(\theta_{\alpha, \bf q_{\parallel}})+\sinh(\theta_{\alpha, \bf q_{\parallel}})$, increases with temperature, from 1 at T=0 to infinity as $T \rightarrow \infty$. The stronger exciton-phonon coupling pushes nuclear equilibrium positions in the excited excitonic state away from the ground-state equilibrium. This, in turn, increases the vertical excitation energy in the Franck-Condon region, shifts the system out of the resonance with the photonic mode, and decreases the polaritonic effects.


Generally speaking, there are two main factors that influence the phonon-assisted transfer between the KK excitons and the intervalley excitons, i.e., the magnitude of the excitation population in the KK state and the phonon number. A higher excitation population or a larger phonon number, which can be achieved {by the lower temperature or the higher temperature, respectively}, facilities the phonon-assisted population transfer. It hints the temperature plays the role of a double-edged sword. For the population transfer channel from the KK state to the KQ state, the magnitude of excitation population of KK state is dominant, also lower frequencies of LA and TA phonons can ensure enough phonon number to assist the intervalley exciton transfer even in lower temperatures. Hence, the population evolution of KQ exciton has a similar temperature-dependent relationship as the KK exciton within 250 fs, as shown in Fig.~\ref{Fig1}(g). Certainly, the exciton transfer between the $\rm{KK^{\prime}}$ state and the KQ state also contributes to the exciton evolution in the KQ state, but this contribution is relatively small in the initial 100 fs due to the small population of $\rm{KK^{\prime}}$ exciton.
Furthermore, the population of the KQ exciton starts showing temperature dependence after roughly 50 fs, while for the KK exciton this happens  after 20 fs [see Figs.~\ref{Fig1}(f) and (g)]. We thus conclude that the effect of temperature on the exciton dynamics is substantial  only if excitonic populations are sufficiently large.


In Fig.~\ref{Fig1}(h), the populations of the $\rm{KK^{\prime}}$ exciton at 75 K and 150 K are in general higher than that at 225 K and 300 K. However, the  $\rm{KK^{\prime}}$ exciton at 300 K has higher population than at 225 K, breaking the monotonic scaling with temperature.
{For the intervalley exciton transfer between the KK state and the $\rm{KK^{\prime}}$ state, on the one hand, the higher-frequency LA and TA phonon modes take part in assisting this transfer, thus a lower temperature may hinder the exciton transfer due to a smaller phonon number. This competitive mechanism brings about a more complex temperature-dependent relationship for the population evolution in the $\rm{KK^{\prime}}$ state. On the other hand, the multiple transfer channels between the $\rm{KK^{\prime}}$ state and the other excitonic states will also lead to the elusive dynamic behaviors.}



\begin{figure*}[htb]
\centering
\subfigure[]{
\includegraphics[scale=0.25,trim=80 0 90 0]{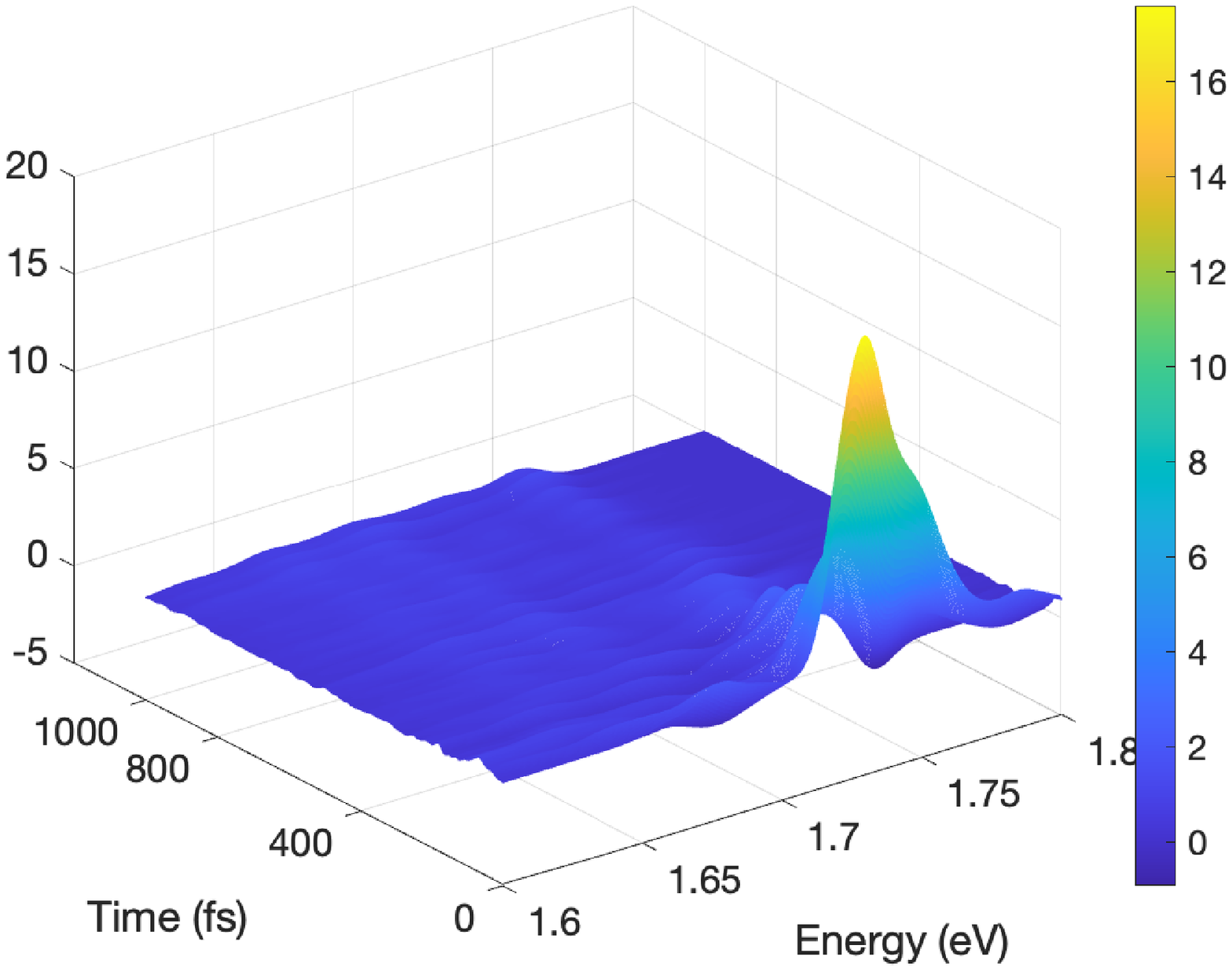}
}
\quad
\subfigure[]{
\includegraphics[scale=0.25,trim=30 0 90 0]{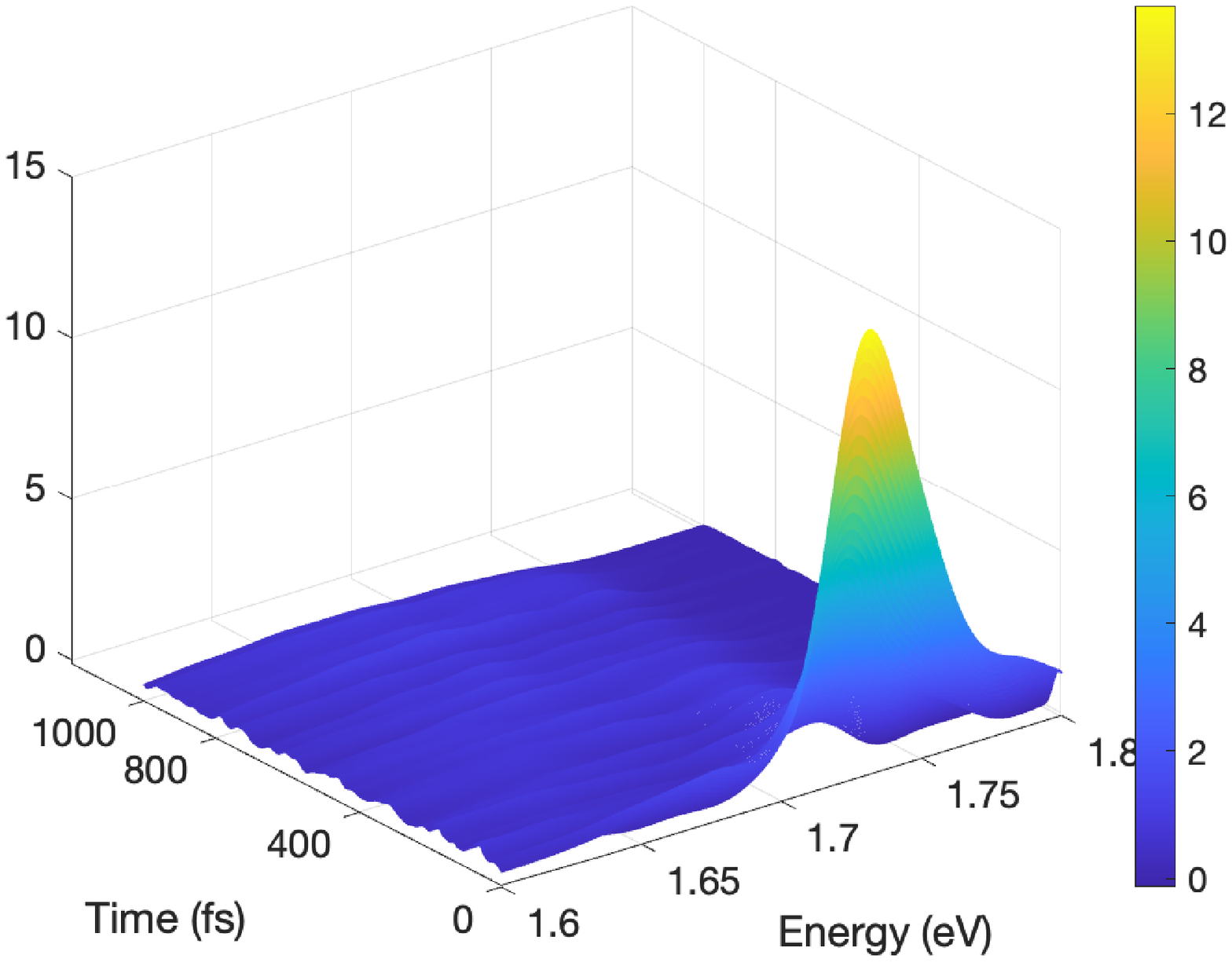}
}
\quad
\subfigure[]{
\includegraphics[scale=0.25,trim=30 0 90 0]{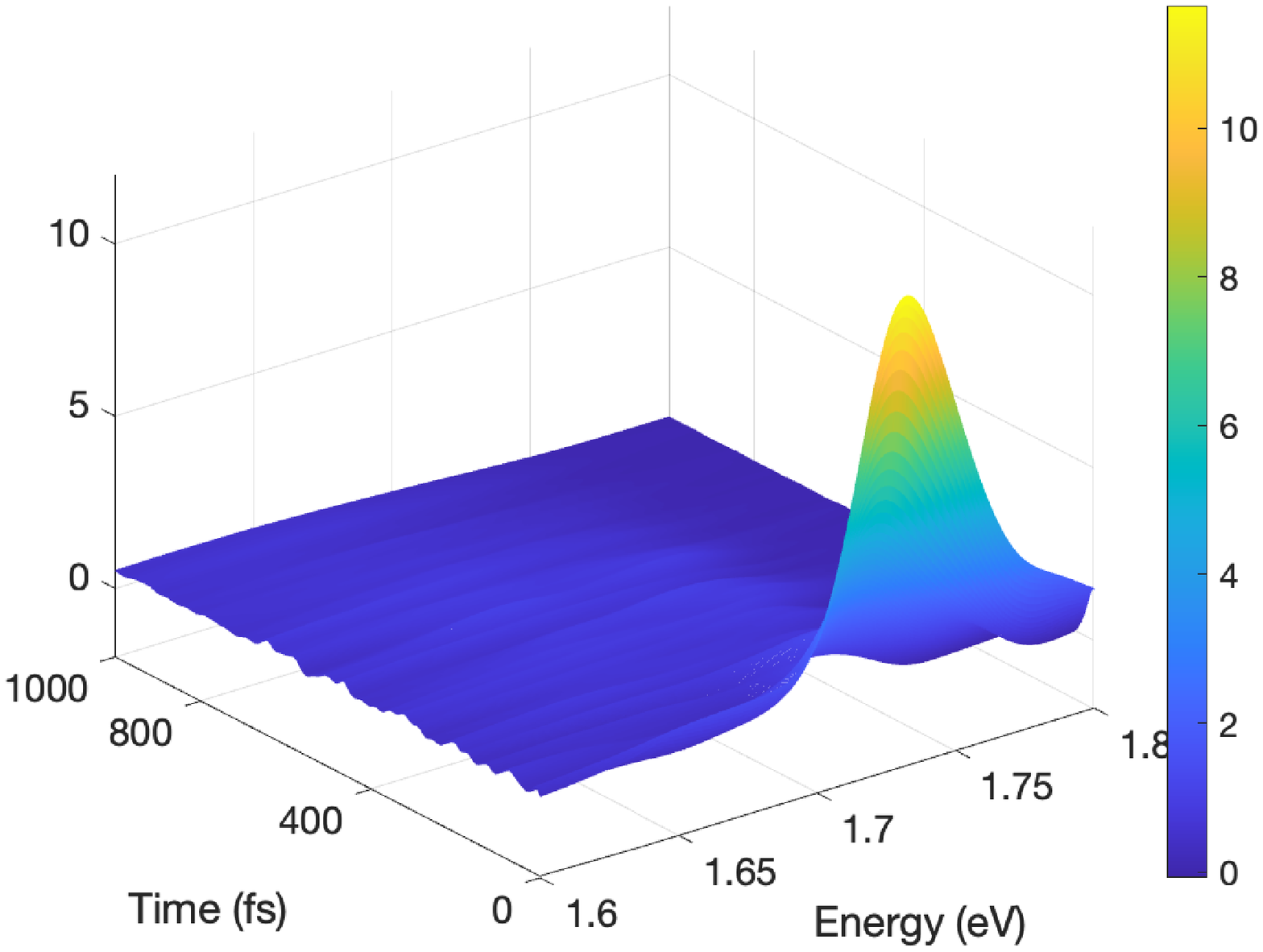}
}
\quad
\subfigure[]{
\includegraphics[scale=0.25,trim=30 0 90 0]{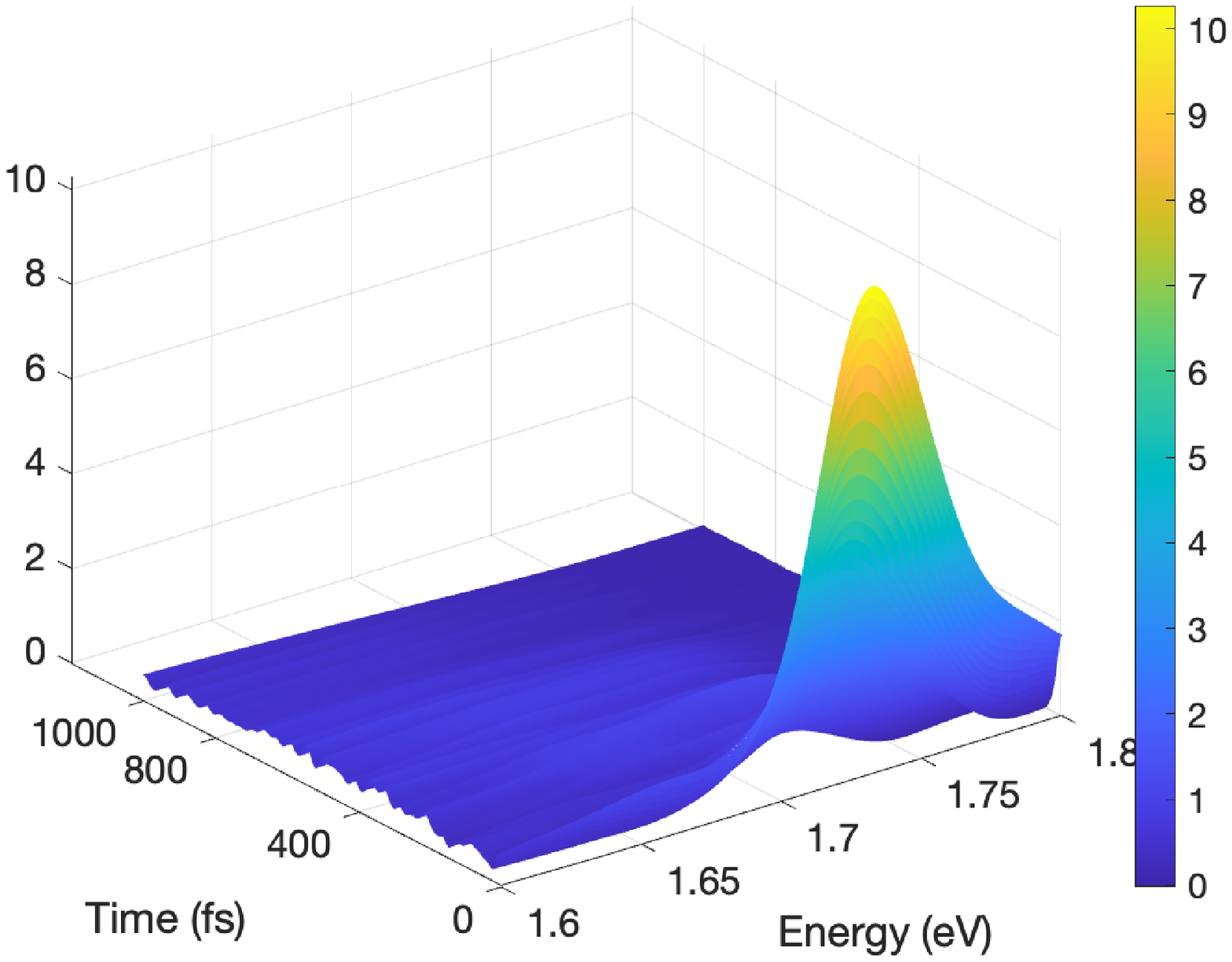}
}\\
\subfigure[]{
\includegraphics[scale=0.25,trim=80 0 80 0]{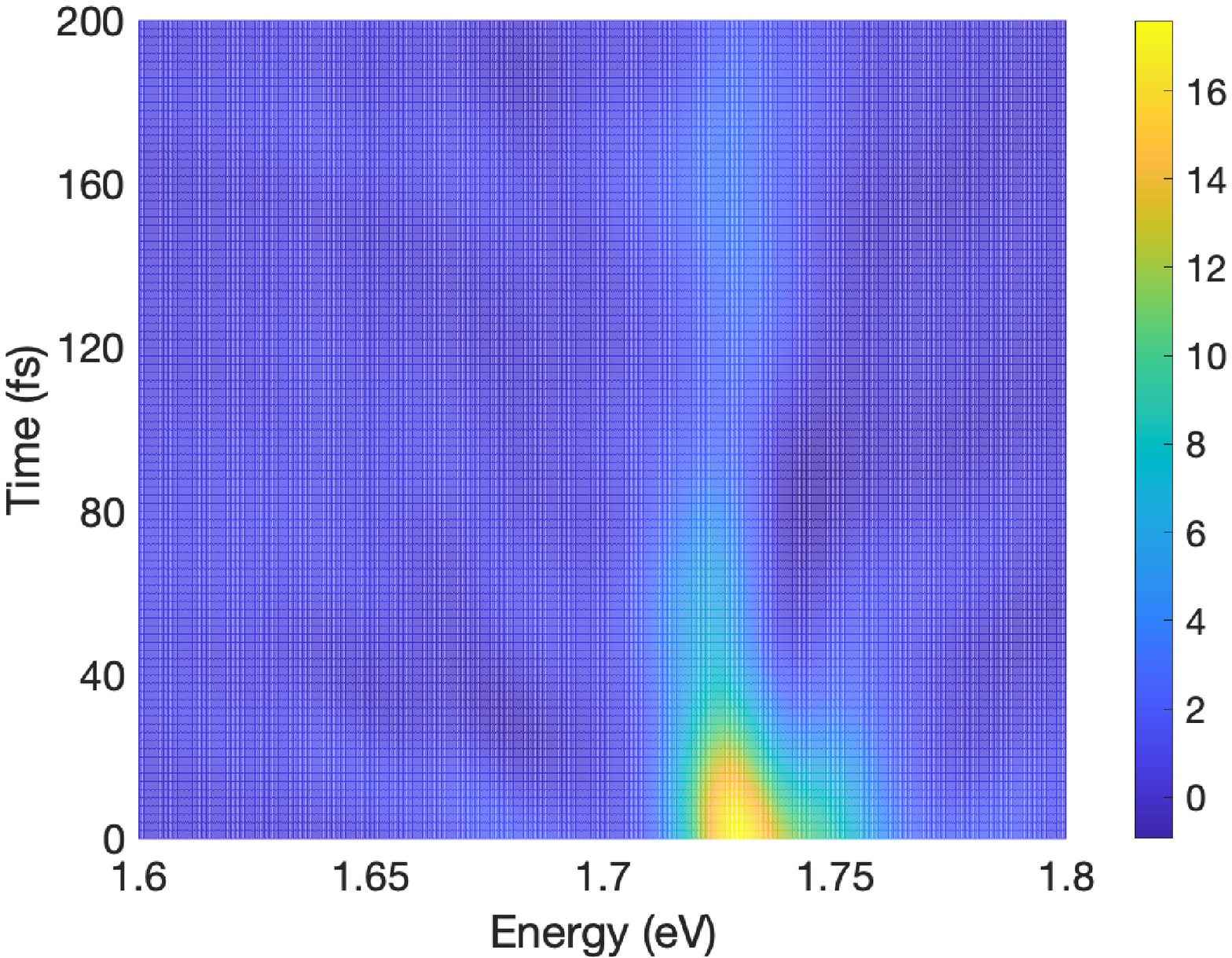}
}
\quad
\subfigure[]{
\includegraphics[scale=0.25,trim=30 0 80 0]{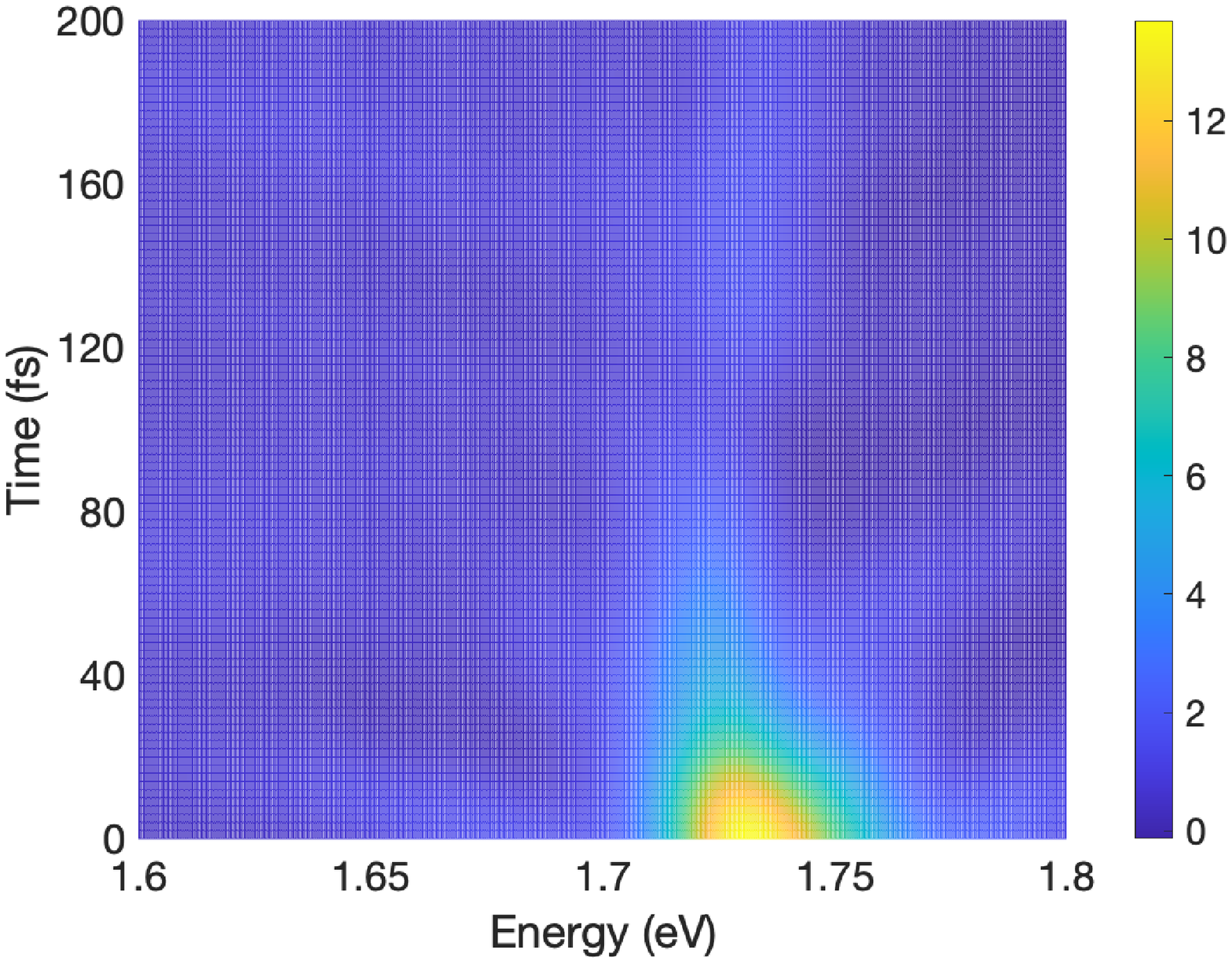}
}
\quad
\subfigure[]{
\includegraphics[scale=0.25,trim=30 0 80 0]{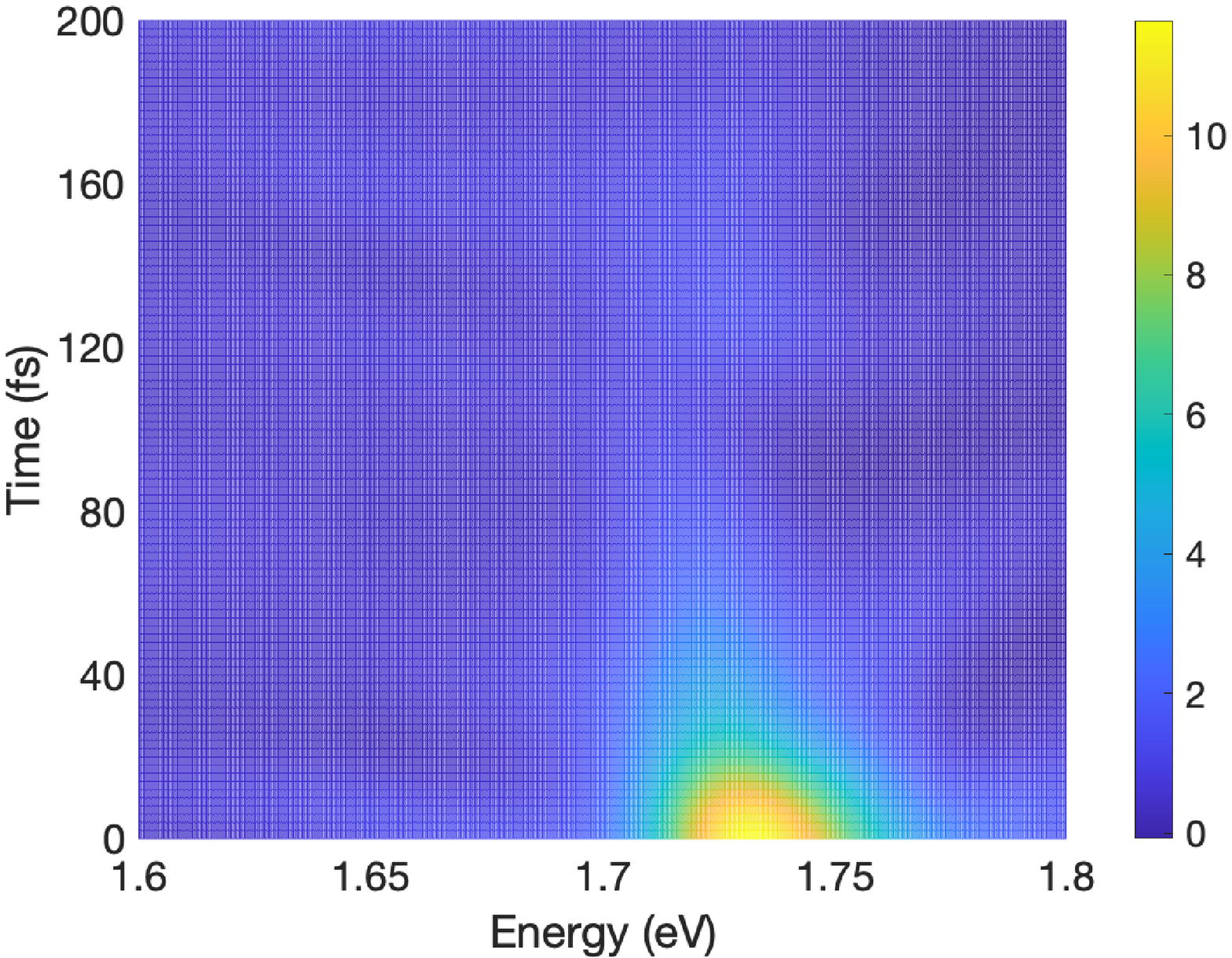}
}
\quad
\subfigure[]{
\includegraphics[scale=0.25,trim=30 0 80 0]{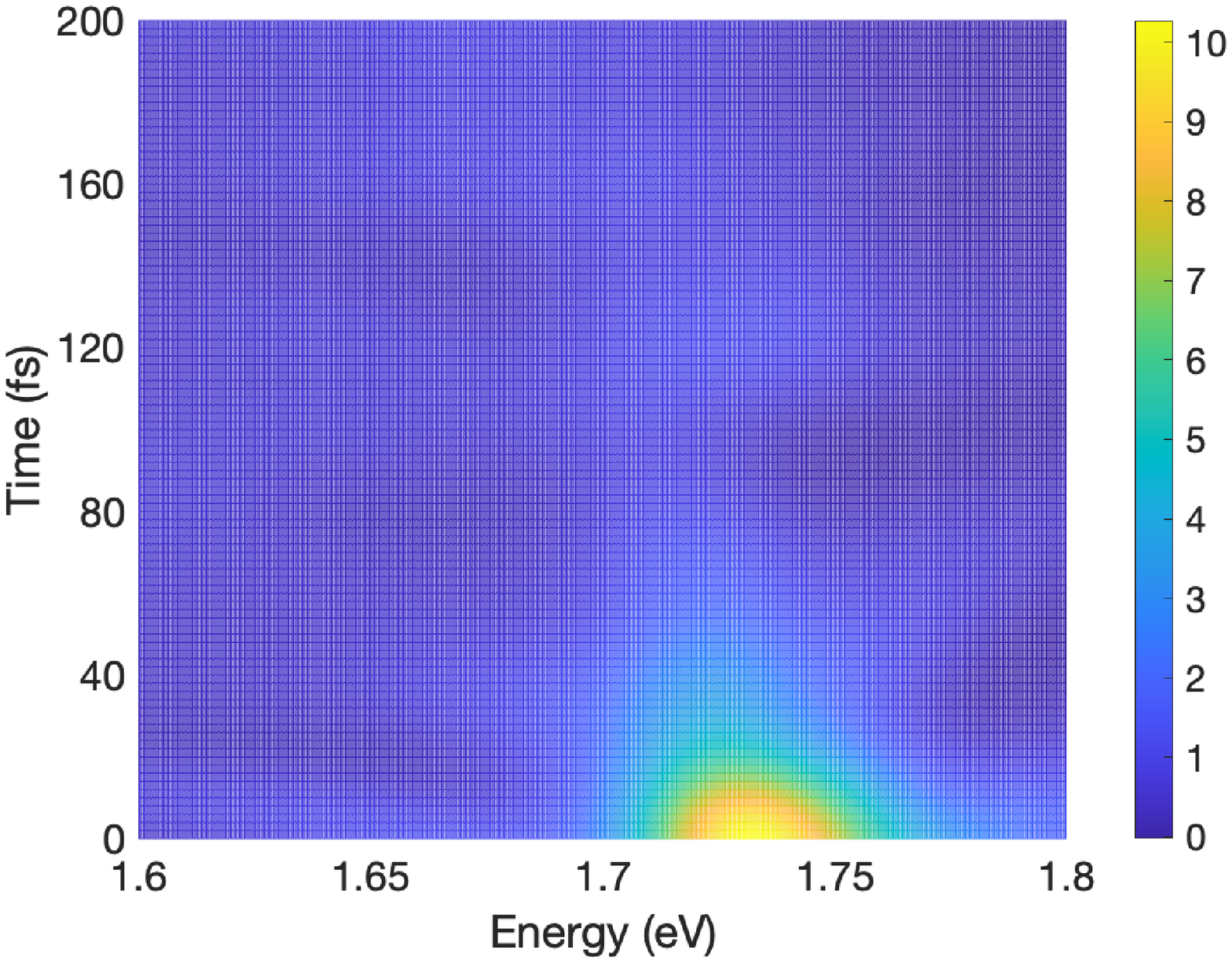}
}\\
\caption{TRF spectra $S(\omega, t)$ of the WSe$_2$ system at  75 K (a, e), 150 K (b, f),  225 K (c, g), and 300 K (d, h).
Upper panels: 3D view for $0<t<1000$ fs. Lower panels: 2D view for $0<t<200$ fs.}
\label{Fig2}
\end{figure*}

For extracting more detailed information on the ultrafast excitonic dynamics in the  WSe$_2$ monolayer, we simulated TRF spectra which -- under assumption of instantaneous excitation of the system by a short pump pulse -- can be  evaluated by the formula \cite{Mukamel,ZYKnox,Gelin02}
\begin{equation}
S(\omega, t) \sim {\rm Re} \int_{0}^{\infty} {\rm d} t_3 R_1^{DA} (t_3,t,0) R_1^{g(t)} (t_3,t,0) {\rm exp} [i\omega t_3] \label{Fl}
\end{equation}
where $R_1^{DA} (t_3,t,0)$ is the third-order response function calculated via the multi-$\rm D_2$ DA and   $R_1^{g(t)} (t_3,t,0)$ is the response function of the environment, which accounts for the TMD degrees of freedom that are not included in the Hamiltonian of Eq. (\ref{H}). $R_1^{g(t)} (t_3,t,0)$ is evaluated  with the lineshape function \cite{Mukamel,ZYKnox}
\begin{align}\label{gt}
	&g(t) = g'(t)+ig''(t) \\
	&g''(t) = -(\lambda / \Lambda)[\rm {exp}(-\Lambda t)+ \Lambda t - 1] \\
	&g'(t) = (\lambda / \Lambda) \rm {cot}(\hbar \beta \Lambda /2)[\rm {exp}(-\Lambda t)+ \Lambda t - 1] \nonumber\\
	& +  \frac {4 \lambda \Lambda} {\hbar \beta} \sum_{n = 1}^{\infty} \frac {\rm {exp}(-\upsilon_n t)+ \upsilon_n t -1} {\upsilon_n({\upsilon_n}^2- \Lambda^2)}\\
	&\upsilon_n = \frac {2 \pi} {\hbar \beta} n
\end{align}
($\lambda$ is the Stokes shift and $\Lambda^{-1}$ is the memory time) which gives shape to the spectral features. The explicit expressions for $R_1^{DA} (t_3,t,0)$ and  $R_1^{g(t)} (t_3,t,0)$ are given in Supporting Information.

Fig.~\ref{Fig2} provides the general view (upper panels) and elucidates the short-time behavior (lower panels) of the TRF signals $S(\omega, t)$ calculated with the lineshape function of Eq. (\ref{gt}) for $\lambda=5$ meV and $\Lambda=3$ meV.
The upper panels reveal fast fluorescence decay of the bright KK exciton (direct PL), which exhibits a $\approx 90\%$ intensity drop within the first 100 fs. This  is a signature of the ultrafast internal conversion at conical intersections~\cite{skw1,skw2,Sun3,ConicalIntersections}. The lower panels zoom into the short-time evolution  of the TRF signals and clarify the inner kitchen of the conical-intersection-driven dynamics. All spectra are grouped around $\omega=1.724$ eV, which corresponds to the bright KK exciton. On the one hand, this proves that the lower-lying KQ and $\rm{KK^{\prime}}$ excitons, being optically dark, do not emit at short times (cf. Ref. \cite{LP_SE}). On the other hand, this is a manifestation of the polaritonic effect: the bare excitonic KK state splits into a pair of bright polaritonic states  separated by, approximately, $2M_{\sigma_{-}}=24~\rm meV$. This causes broadening of the spectrum. Coupling to the phonon modes (polaron dressing)  also broadens the spectrum, elongating it towards the blue wing. Furthermore, the phonon-assisted processes  are responsible for the significant temperature dependence of the spectra: as temperature increases,  $S(\omega, t)$  broaden  along the $\omega$-axis, become more symmetric, and depopulate faster. The latter effect is generic for conical intersections produced by low-frequency coupling modes \cite{skw20}.

\begin{figure*}[tbp]
\centering
\subfigure[]{
\includegraphics[scale=0.25,trim=80 0 80 0]{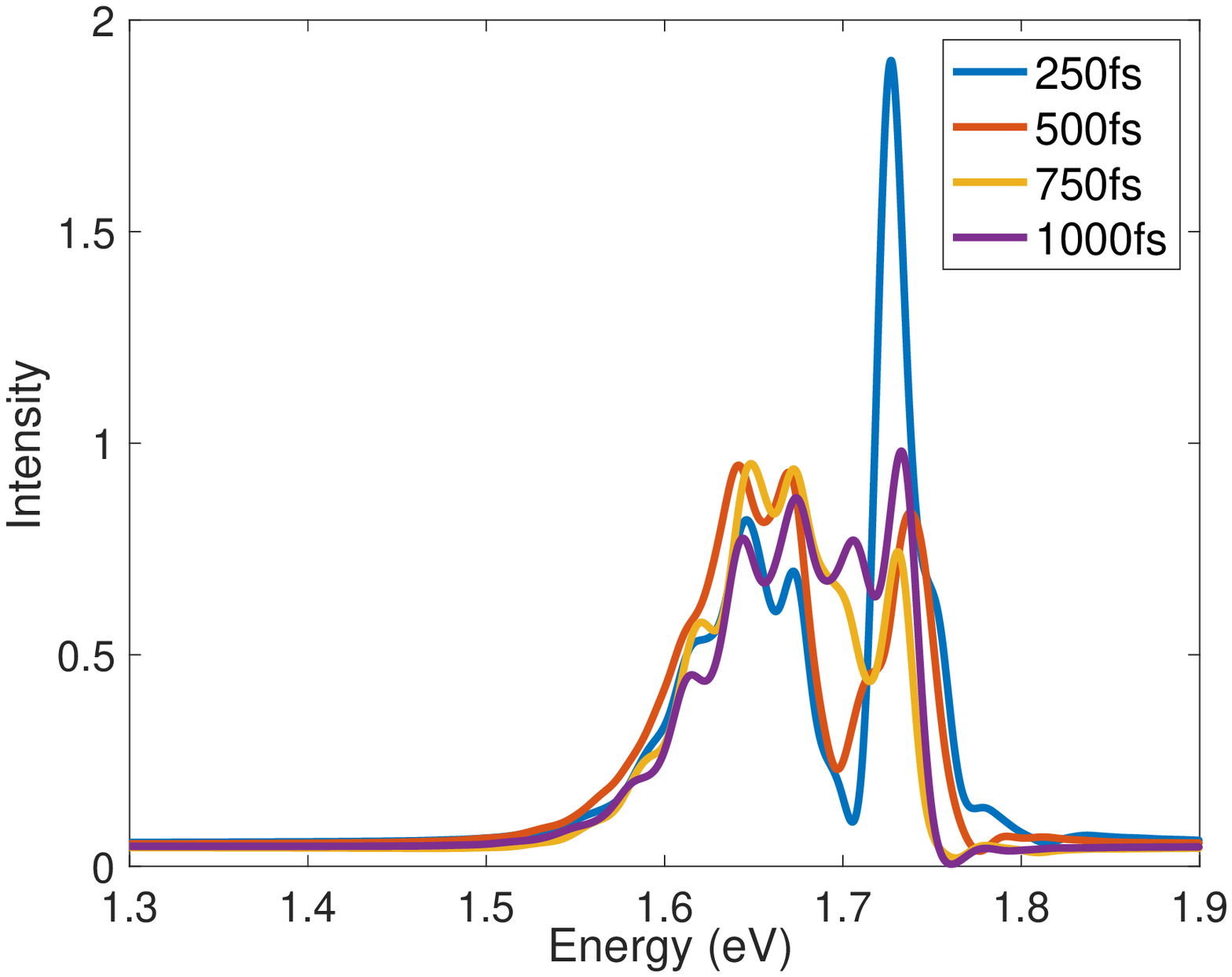}
}
\quad
\subfigure[]{
\includegraphics[scale=0.25,trim=30 0 80 0]{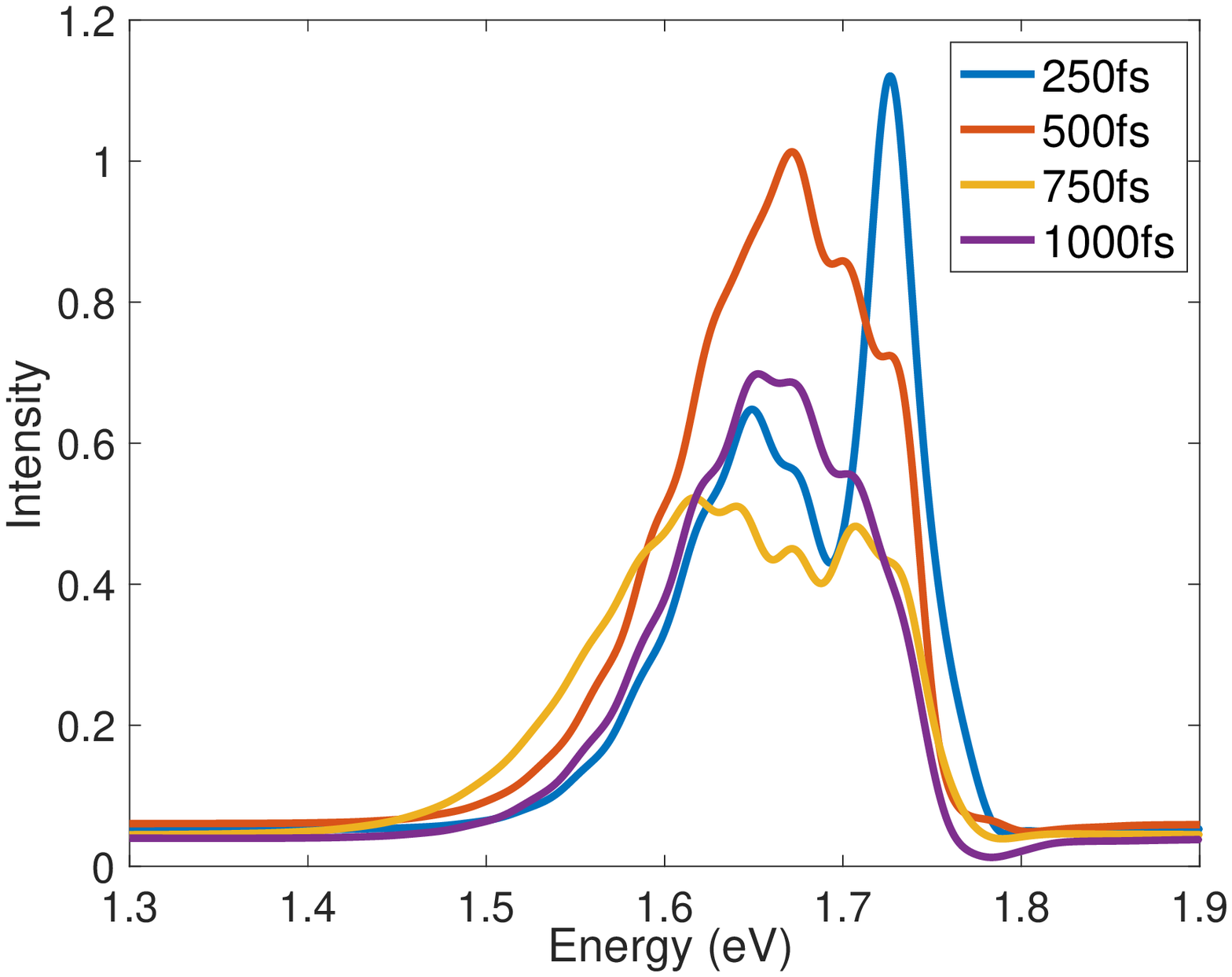}
}
\quad
\subfigure[]{
\includegraphics[scale=0.25,trim=30 0 80 0]{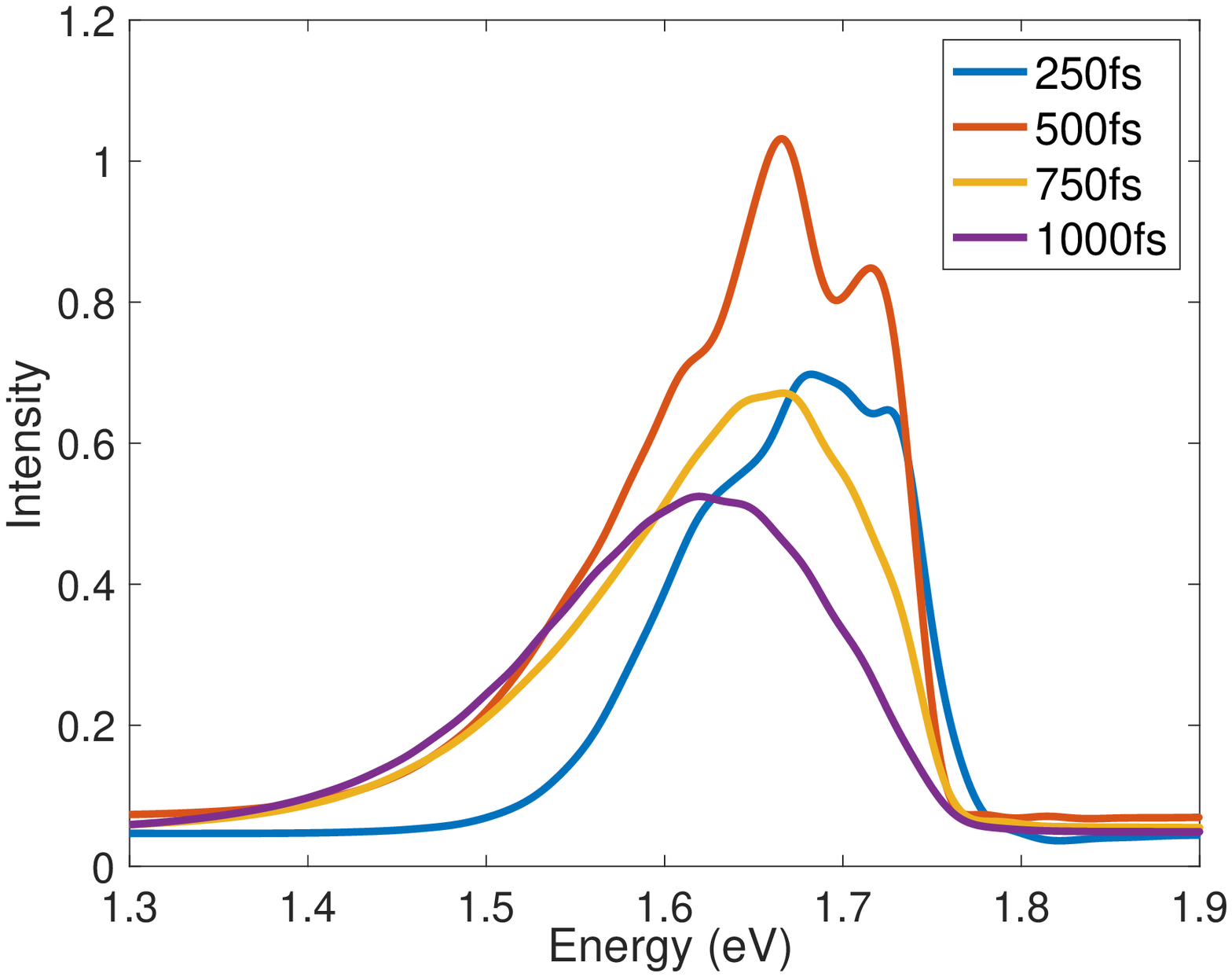}
}
\quad
\subfigure[]{
\includegraphics[scale=0.25,trim=30 0 80 0]{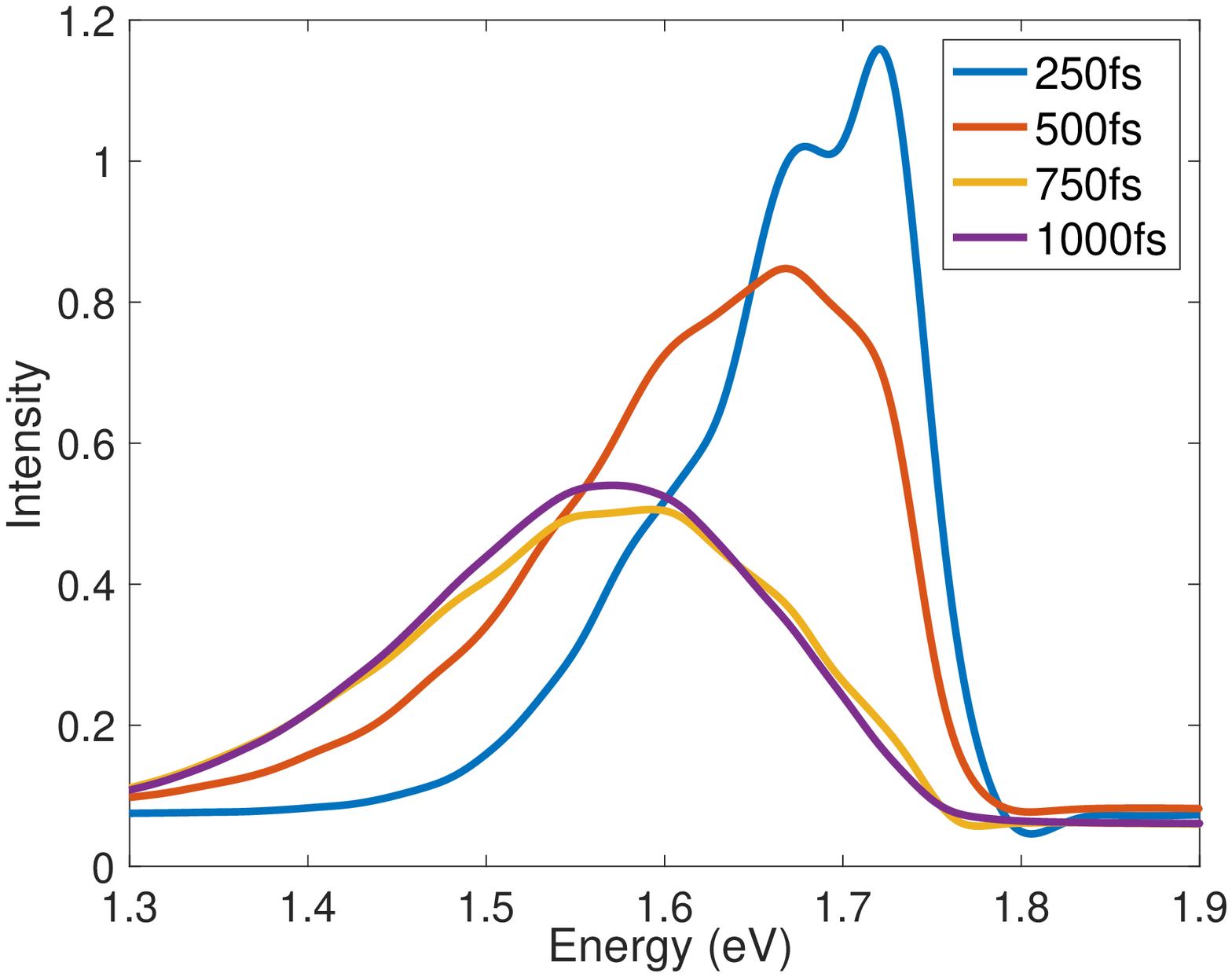}
}\\
\subfigure[]{
\includegraphics[scale=0.25,trim=80 0 80 0]{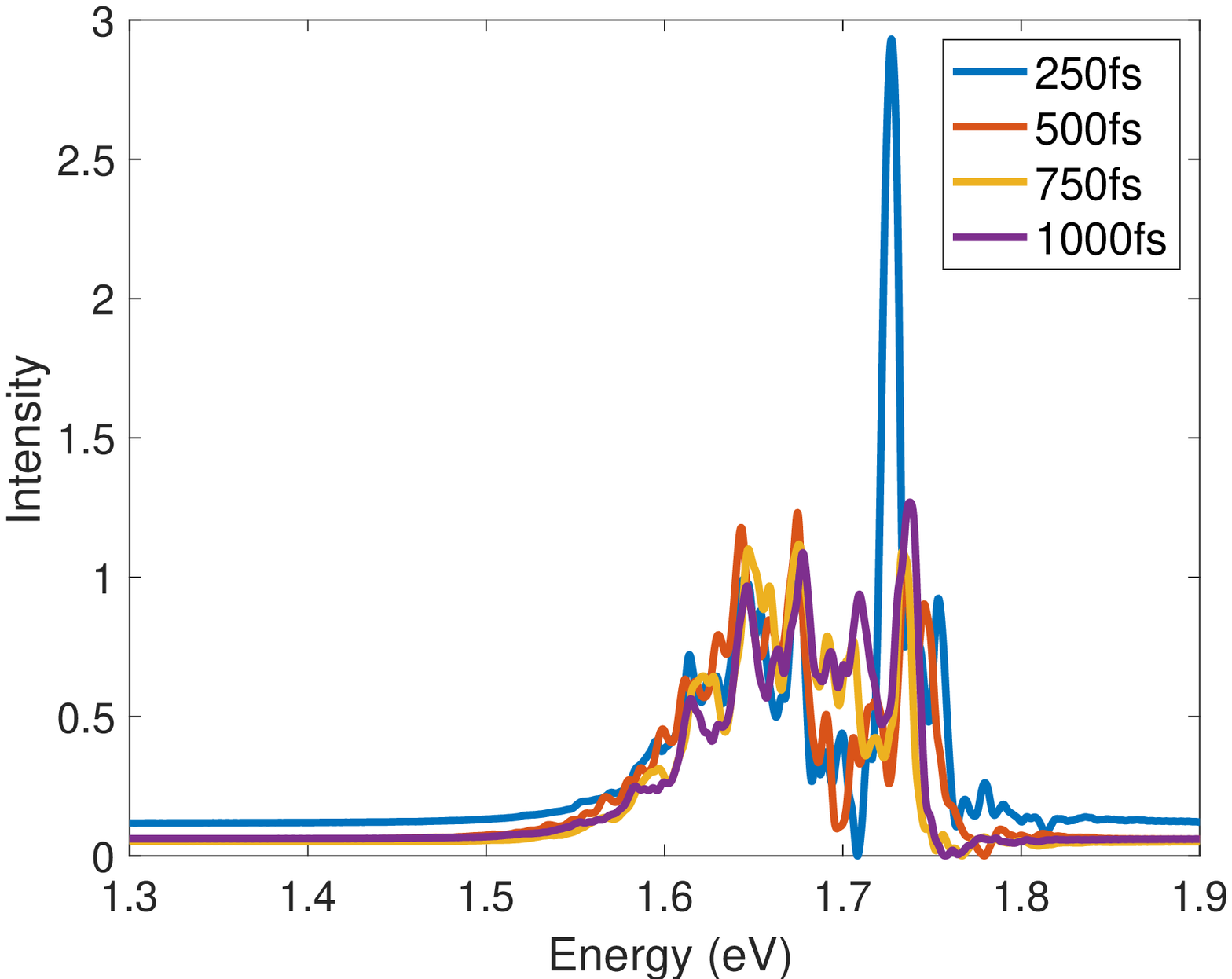}
}
\quad
\subfigure[]{
\includegraphics[scale=0.25,trim=30 0 80 0]{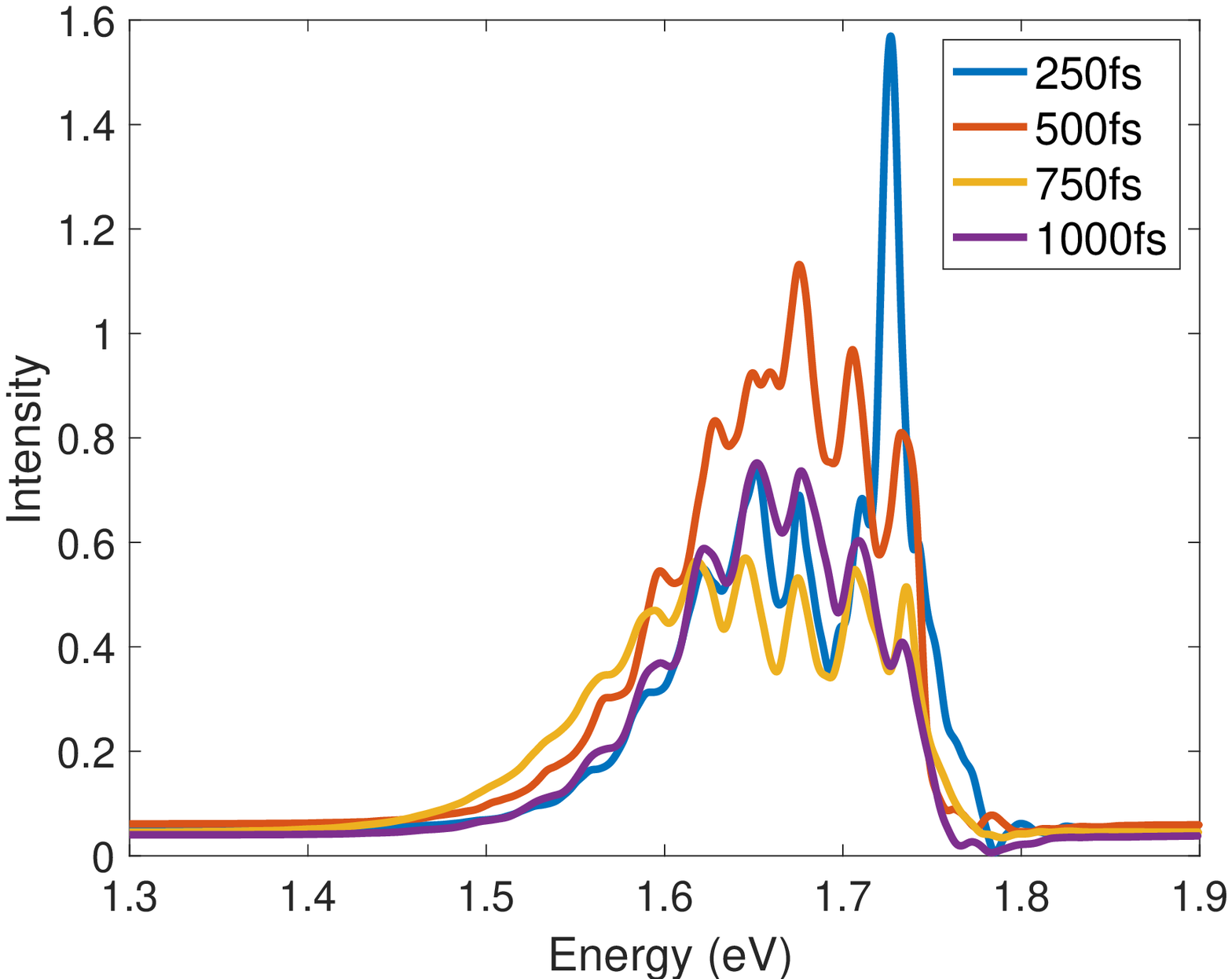}
}
\quad
\subfigure[]{
\includegraphics[scale=0.25,trim=30 0 80 0]{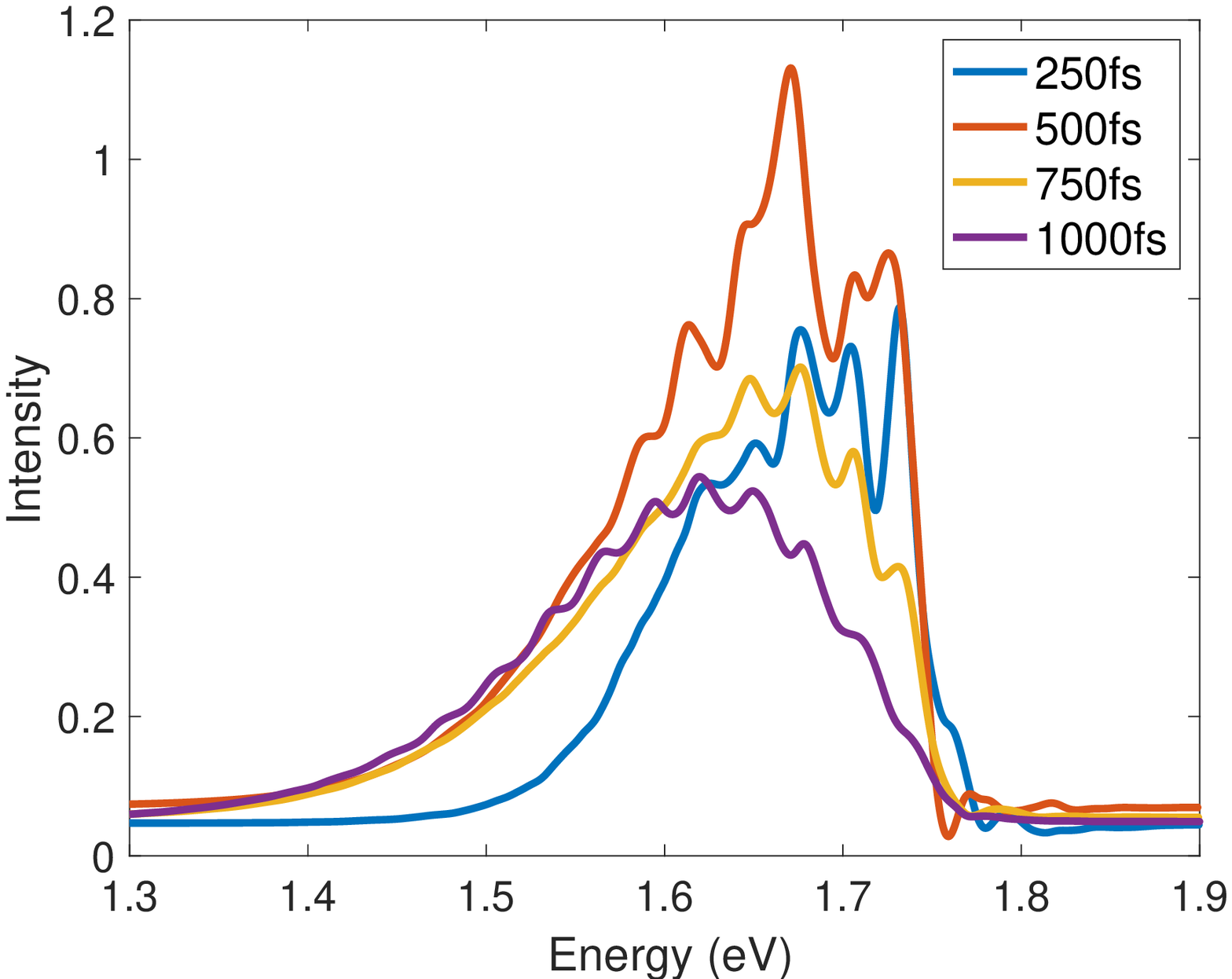}
}
\quad
\subfigure[]{
\includegraphics[scale=0.25,trim=30 0 80 0]{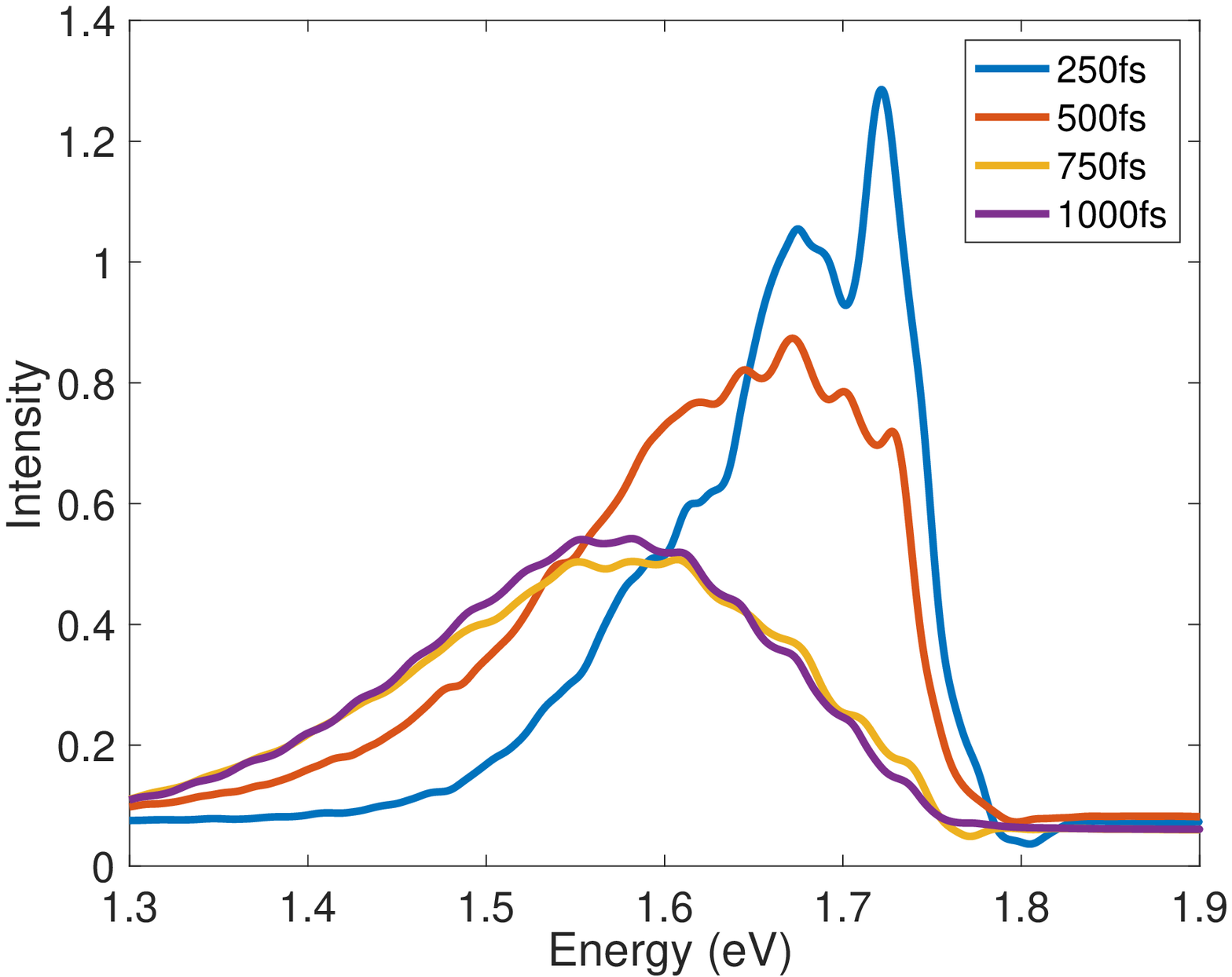}
}\\
\caption{Fluorescence spectra $S(\omega, t)$ of the WSe$_2$ system for $t=250$, 500, 750, and 1000 fs at  75 K (a, e), 150 K (b, f),  225 K (c, g), and 300 K (d, h).
	Upper panels: $\lambda=5$ meV and $\Lambda=3$  meV. Lower panels: $\lambda=5$ meV and $\Lambda=30$  meV. }
\label{Fig3}
\end{figure*}

Fig.~\ref{Fig3} provides a detailed view of the TRF spectra at longer time, showing the fluorescence profiles $S(\omega, t)$ at $t=250$, 500, 750, and 1000 fs for increasing temperatures (from left to right, $T=75$, 150, 225 and 300 K) and two sets of parameters specifying the lineshape function of Eq. (\ref{gt}). In the upper panels, the Stokes shift is fixed at $\lambda = 5$  meV and $\Lambda = 3$ meV (relatively slow spectral diffusion). In the lower panels,  $\lambda = 5$  meV  and $\Lambda = 30$ (relatively fast spectral diffusion).

We start from a general description. In contrast to Fig.~\ref{Fig2}, the spectra in Fig.~\ref{Fig3} are much broader, extending from $\omega \approx 1.5$ eV in the red to $\omega \approx 1.8$ eV in the blue. Not surprisingly, $S(\omega, t)$ at low and high temperatures are qualitatively different.  At low temperatures (two leftmost columns in Fig.~\ref{Fig3}), the spectra exhibit a clear multi-peak structure, featuring the bright KK state (direct PL) as well as the dark KQ and $\rm{KK^{\prime}}$ states (indirect PL). It is essential that the number of the PL peaks exceeds 4, revealing vibronic features of the TRF resonances. In addition, ``centers of mass" of the spectra do not substantially shift with time, so that $S(\omega, t)$ at different $t$ merely exhibit variations of the peak intensities. This is  a manifestation of the fact that the Hilbert space spanned by the vibrational subsystem is relatively low-dimensional,
so that the phonon modes at low temperatures cannot be considered as a true thermal bath.
 At higher temperatures (two rightmost columns in Fig.~\ref{Fig3}) the spectral features  merge. At $t=250$ fs, $S(\omega, t)$ exhibits highly asymmetric spectral profile elongated to the red, featuring the KK state. At longer times, $S(\omega, t)$ becomes broader, more symmetric and almost featureless, their intensities decrease while their centers of mass move to the red. This indicates that the Hilbert space spanned by the vibrational subsystem becomes substantial and the phonon bath drives the phonon-dressed excitonic system to the equilibrium at lower energies.

Let us now focus on the finer details of the spectral evolution. The upper panels of Fig.~\ref{Fig3} correspond to the relatively large electronic dephasing (which is inversely proportional to $\Lambda$ in the classical limit of $\beta\Lambda \ll 1$, see \cite{Mukamel,ZYKnox}) and relatively slow spectral diffusion ($\Lambda^{-1}=220$ fs). At low temperature and $t=250$ fs [panel (a)], intensity of the KK peak is twice as high as those of the lower-energy peaks, which arise owing to phonon-assisted processes and  intensity borrowing from the bright KK resonance. At higher temperatures and $t=250$,  the peaks merge [panels (c) and (d)] and produce the asymmetric stretched-to-the-red peak with a pronounced shoulder and maximum around the KK resonance. This happens because the peaks revealing indirect transitions  are dissolved in the main KK resonance,  as the effective bandgap narrows and the energy distribution widens. The spectra at $t=500$ fs are peculiar, notably in panel (c): the surface areas under these TRF profiles are larger than surface areas under $S(\omega, t)$ for other $t$. This is a clear signature of the wave-packet motion, which travels between the bright and dark excitonic and photonic states. Note that the TRF profiles at 750 fs and 1000 fs in panel (d) are nearly the same, indicating that the steady-state fluorescence regime is achieved faster at higher temperatures.
The lower panels in Fig.~\ref{Fig3} show the TRF spectra corresponding to weaker electronic dephasing and fast spectral diffusion ($\Lambda^{-1}=22$ fs). Since the peak broadening is smaller, the spectra reveal narrower peaks and richer peak structures. Apart from that, all general features established for the TRF spectra in the upper panels remain unchanged. Qualitatively, the spectra in the lower panels are similar to those simulated in Ref.~\cite{Ross} by adopting the Lindblad master equation for  a similar TMD-cavity system. However, the inter-peak separations and the widths of our spectra are much larger than those in Ref~\cite{Ross}, owing to the stronger exciton-phonon and exciton-photon coupling adopted in our model. The widths of the spectra in the upper and the lower panel are approximately the same (cf.~Ref.~\cite{Gel}), because -- for the chosen values of $\Lambda$ -- the spectral widths are mostly determined by the area filled by the peaks rather than by the broadening of individual peaks.

\begin{figure*}[tbp]
\centering
\subfigure[]{
\includegraphics[scale=0.3,trim=30 0 30 0]{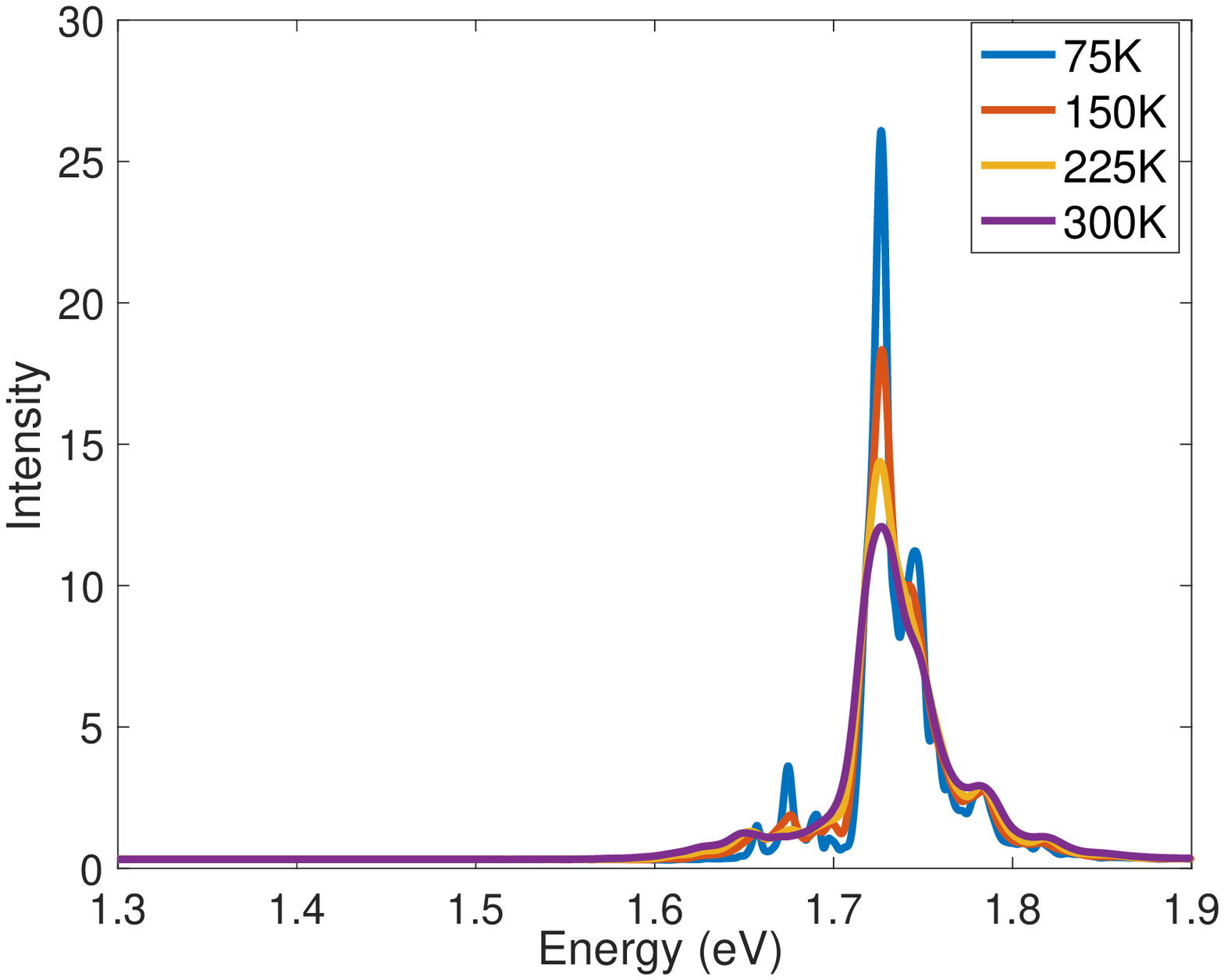}
}
\quad
\subfigure[]{
\includegraphics[scale=0.3,trim=30 0 30 0]{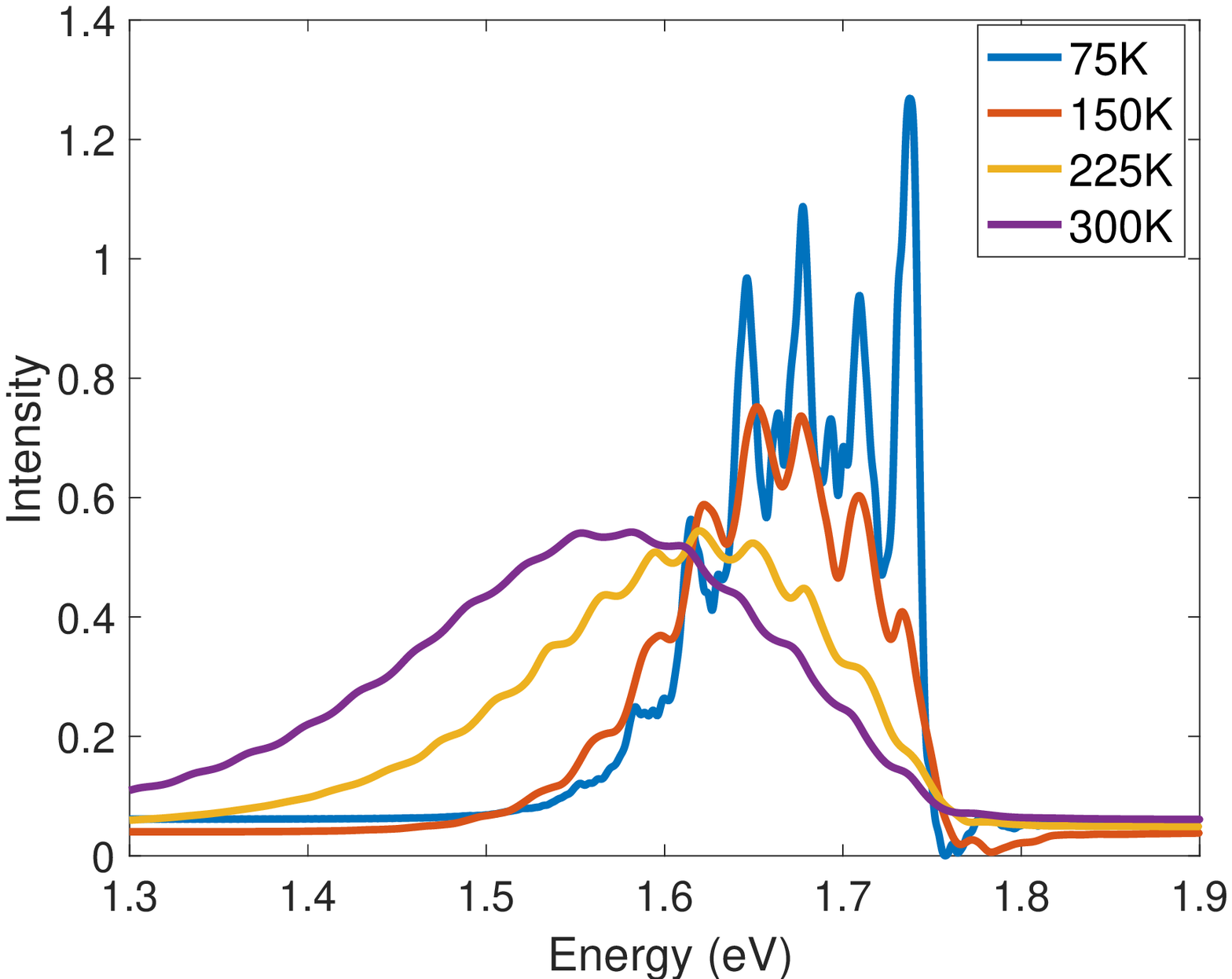}
}
\quad
\subfigure[]{
\includegraphics[scale=0.3,trim=30 0 30 0]{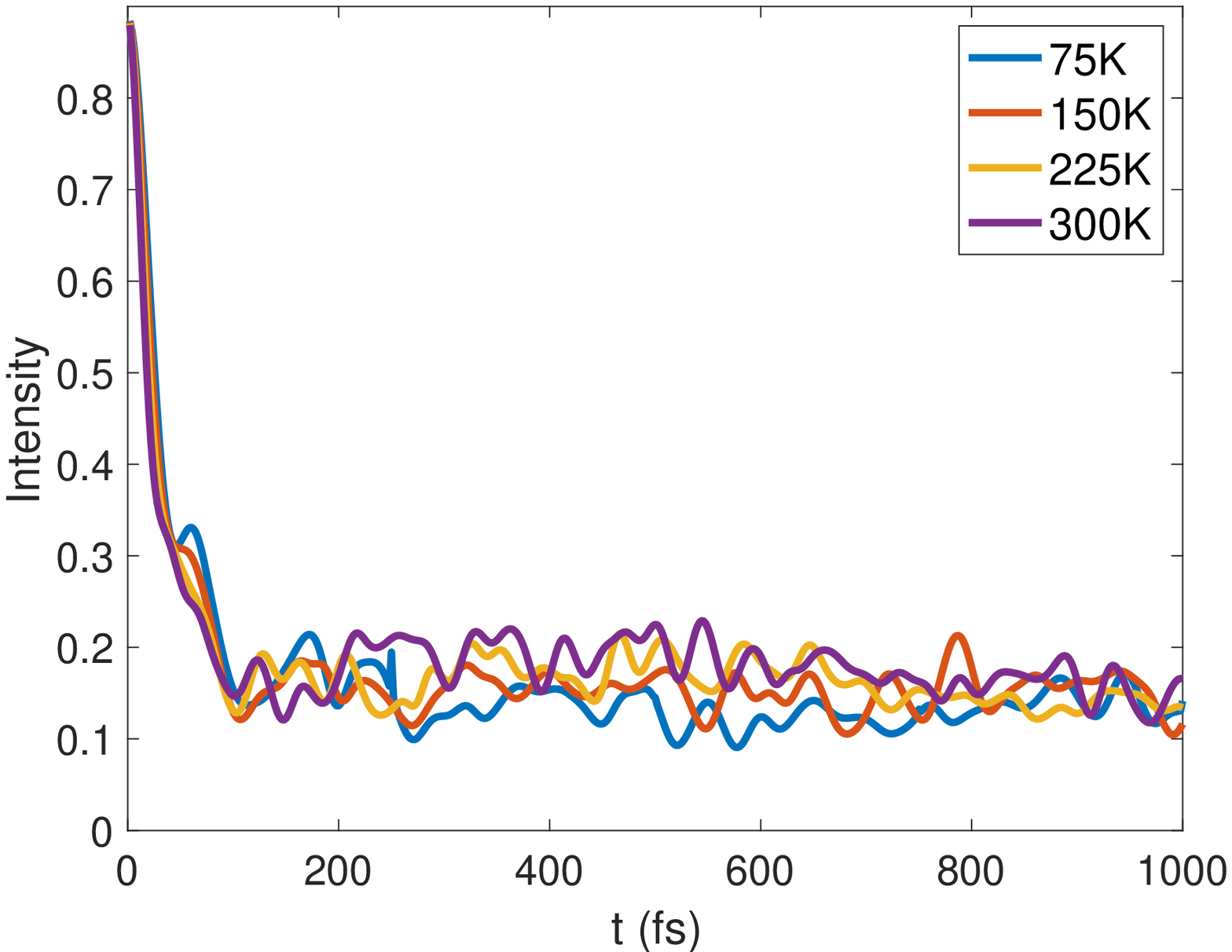}
}\\
\caption{ Fluorescence spectra $S(\omega, t)$ of the WSe$_2$ system for $t=0$ (a) and 1000 fs (b) as well as the integrated TRF signal  $S(t)$ (c) for four temperatures indicated in the panels. $\lambda=5$ meV and $\Lambda=30$ meV.}
\label{Fig6}
\end{figure*}

Fig.~\ref{Fig6} compares profiles of the hot [$t=0$, panel(a)] and relaxed [$t=1000$ fs, panel(b)] TRF spectra of the WSe$_2$ system at four temperatures, while the integrated TRF signal $S(t)=\int d \omega S(\omega, t)$ in panel (c) illustrates the overall intensity change of the spectra in panels (a) and (b). Several hot-fluorescence features were already discussed in the context of Fig.~\ref{Fig2}. Fig.~\ref{Fig6}(a) is intended to explore the fine structure of the spectra  which are hidden in the 2D plots of Fig.~\ref{Fig2}(e)-(h). The low-temperature TRF profile in Fig.~\ref{Fig6}(a) (blue line) reveals three peaks at 1.674, 1.726, and 1.746 eV. The first two peaks correspond to the bare $\rm{KK^{\prime}}$ and KK excitons, correspondingly, while the rightmost peak is separated by $\sim 20$ meV  from the  KK peak and is produced by the photon mode. This reveals that the polaritonic effects are negligible for the $\rm{KK^{\prime}}$ and KK peaks at short times,  because the photon mode needs some time to couple the molecular state produced upon optical excitation of the WSe$_2$ system. On the other hand, the phonon dressing cannot be neglected: if we omit polaronic effects and put $D^{ij}_{\alpha \bf q_{\parallel}}=0$, the two rightmost peaks will be located at 1.712 and 1.736 eV. This is a striking demonstration of the strong entanglement of the electronic and the vibrational degrees of freedom at conical intersections \cite{DS}. Such a strong exciton-phonon coupling strength stabilizes the  KK and $\rm{KK^{\prime}}$  peaks at their bare exciton positions and shifts the polaritonic peak to the blue. Elevated temperatures smear the fine peak structure, broaden TRF profiles and decrease their intensities [see the spectra for T=150, 225, and 300 K in Fig.~\ref{Fig6}(a)].
The relaxed TRF spectra in Fig.~\ref{Fig6}(b) show a qualitatively different picture. Here the low-temperature spectrum (blue line) exhibits a multi-peak progression in which none of the peaks corresponds to the bare excitonic states, emphasizing the significance of the polaron and the polariton effects at longer times even at low temperatures. Elevated temperatures enhance  the impact of these effects [cf.~the spectra at T=150, 225, and 300 K in Fig.~\ref{Fig6}(b)].
The total TRF signal $S(t)$ displayed in Fig.~\ref{Fig6} (c) exhibits two notable features which are typical for conical-intersection systems  \cite{DS}. At short times, it shows a fast, quasi-Gaussian decay. At longer times, it reveals complex oscillatory patterns which are grouped around $I(t) \approx 0.14$ and depend significantly on the  temperature.

In summary, we constructed an ab-initio-parameterized Hamiltonian for a cavity-controlled, single-layered WSe$_2$ system, and combined the multi-$\rm D_{2}$ DA method~\cite{Zhao} with the TFD machinery~\cite{Chen,Borrelli} to accurately simulate the exciton dynamics and the TRF spectra of the many-body, multi-species system at temperatures from 75 K (where the thermal effects are insignificant even for the phonon modes with lowest frequencies) to 300 K (where the thermal effects are essential even for the phonon modes with highest frequencies). This allowed us to establish dynamical and spectroscopic signatures of the polaronic and polaritonic effects at different temperatures as well as to uncover their characteristic timescales. In particular, our studies revealed the pivotal role of the multidimensional conical intersections in controlling dynamics of the strongly coupled excitonic, phononic, and photonic modes.

Technically, the WSe$_2$ Hamiltonian describes three excitonic states, whose potential energy surfaces cross each other via multidimensional conical intersections shaped by 23 coupling modes. To our knowledge, this is the most complex multidimensional conical intersection system, the dynamics and spectroscopic responses of which have been studied by a numerically accurate, fully quantum method. Hence, the computationally efficient method employed here can be recommended for future ab-initio-based simulations of cavity-controlled 2D materials  probed by various nonlinear spectroscopic techniques at finite temperatures.

\section*{Acknowledgments}
We would like to thank Fulu Zheng for assistance with computation at the last stage of the work. The authors
gratefully acknowledge the support of the Singapore Ministry of Education Academic Research Fund (Grant Nos.~RG190/18 and RG87/20). K.~Sun would also like to thank the Natural Science Foundation of Zhejiang Province (Grant No.~LY18A040005) for partial support. M. F. G. acknowledges the support of Hangzhou Dianzi University through startup funding.
\section*{Data Availability}
The data that support the findings of this study are available from the corresponding author upon reasonable request.

\end{document}